\documentclass[12pt]{article}
\usepackage{multirow}
\usepackage{graphics}
\usepackage{lscape}
\usepackage{amsfonts}
\usepackage{amsmath}
\usepackage{amssymb}
\usepackage{rotating}
\usepackage{amssymb,amsmath}
\usepackage[usenames]{color}
\usepackage{graphicx}
\usepackage[english]{babel}
\usepackage{natbib}
\usepackage{array}
\newcolumntype{P}[1]{>{\centering\arraybackslash}p{#1}}
\usepackage{array,multirow}
\usepackage{hhline}

\usepackage{dcolumn}
\newcolumntype{d}[1]{D{.}{.}{#1}}
\usepackage{float}
\usepackage{xcolor}
\usepackage{ulem} % This makes the text strike out.
\usepackage{actuarialangle}
\usepackage[flushleft]{threeparttable}
\usepackage{comment}
\makeatletter

\newcommand{\Rmnum}[1]{\expandafter\@slowromancap\romannumeral #1@}
\makeatother
\usepackage{etoolbox}
\appto\TPTnoteSettings{\footnotesize}
\renewcommand{\section}[1]{\refstepcounter{section}\begin{center}{\sc \thesection. #1}\end{center}}
\renewcommand{\subsection}[1]{\bigskip \noindent \refstepcounter{subsection}{\thesubsection\ {\it #1}}}
\renewcommand{\subsubsection}[1]{\bigskip \noindent \refstepcounter{subsubsection}{\thesubsubsection\ {\it #1}}}
\renewcommand{\title}[1]{{\bf \begin{center}#1\end{center}}}
\renewcommand{\author}[1]{\begin{center}{\sc #1}\end{center}}

\newcommand{\acknowledgements}{\begin{center}{\sc Acknowledgements}\end{center}}
\newcommand{\competinginterests}{\begin{center}{\sc Competing Interests}\end{center}}
\renewenvironment{abstract}{\begin{center}{\sc abstract}\end{center}\small}{\normalsize}
\newenvironment{keywords}{\begin{center}{\sc keywords}\end{center}\small}{\normalsize}

\newenvironment{contacts}{\begin{center}{\sc contact addresses}\end{center}\small}{\normalsize}

\newenvironment{bajlist}{\begin{list}{(\alph{bajlistnum})\hfill}{\usecounter{bajlistnum}\setlength{\labelwidth}{0.3in}\setlength{\leftmargin}{0.3in}\setlength{\rightmargin}{0in}\setlength{\labelsep}{0in}\setlength{\topsep}{0in}\setlength{\partopsep}{0in}\setlength{\itemsep}{0in}\setlength{\parsep}{0in}}}{\end{list}{}\bigskip}
\newenvironment{bajsublist}{\begin{list}{(\arabic{bajsublistnum})\hfill}{\usecounter{bajsublistnum}\setlength{\labelwidth}{0.3in}\setlength{\leftmargin}{0.3in}\setlength{\rightmargin}{0in}\setlength{\labelsep}{0in}\setlength{\topsep}{0in}\setlength{\partopsep}{0in}\setlength{\itemsep}{0in}\setlength{\parsep}{0in}}}{\end{list}{}}

\newcommand{\Prob}{\mbox{P}}
\newcommand{\Mean}{\mbox{E}}
\newcommand{\Var}{\mbox{Var}}

\newcounter{bajlistnum}
\newcounter{bajsublistnum}

\newtheorem{proposition}{Proposition}
\newtheorem{corollary}{Corollary}

\begin{document}

\newcommand{\HRule}{\rule{\linewidth}{0.5mm}} % Defines a new command for the horizontal lines, change thickness here

%----------------------------------------------------------------------------------------
%	TITLE SECTION
%----------------------------------------------------------------------------------------

\title{LAPSE-SUPPORTED LIFE INSURANCE AND ADVERSE SELECTION}

\author{By Oytun Ha\c{c}ar{\i}z\dag, Torsten Kleinow\ddag $\,$ and  Angus S. Macdonald\S}

\begin{abstract}

\noindent If individuals at the highest mortality risk are also least likely to lapse a life insurance policy, then lapse-supported premiums magnify adverse selection costs. As an example, we model `Term to 100' contracts, {and risk as revealed by genetic test results}. We identify three methods of managing lapse surplus: eliminating it by design; disposing of it retrospectively (through participation); or disposing of it prospectively (through lapse-supported premiums). We then assume a heterogeneous population in which: (a) insurers cannot identify  individuals at high mortality risk; (b) a secondary market exists that prevents high-risk policies from lapsing; (c) financial underwriting is lax or absent; and (d) life insurance policies may even be initiated by third parties as a financial investment (STOLI). Adverse selection losses under (a) {are typically very small}, but under (b) can be increased by multiples, and under (c) and (d) increased almost without limit. We note that the different approaches to modeling lapses used in studies of adverse selection and genetic testing appear to be broadly equivalent and robust.

% All three give broadly similar mortality losses, greatly magnified by a deficiency of lapses among higher risks, but only (c) is sensitive to the lapse assumption in the premium/valuation basis, and this is the distinctive risk under lapse-supported premiums {TO BE REDRAFTED}.

\end{abstract}


\begin{contacts}

\noindent \dag $\,$ Department of Actuarial Sciences, The Faculty of Business, Karab{\"{u}}k University, Karab{\"{u}}k, 78050, Turkey, and Institute of Applied Mathematics, Middle East Technical University, Ankara, 06800, Turkey.

\noindent \ddag $\,$ Research Centre for Longevity Risk, Faculty of Economics and Business, University of Amsterdam.

\noindent \S $\,$ Department of Actuarial Mathematics and Statistics, Heriot-Watt University, Edinburgh EH14 4AS, UK, and the Maxwell Institute for Mathematical Sciences, UK.

\noindent Corresponding author: Angus Macdonald, A.S.Macdonald@hw.ac.uk.

\end{contacts}

%----------------------------------------------------------------------------------------
%	END TITLE SECTION
%----------------------------------------------------------------------------------------

\section{Introduction}
\label{sec:Intro}

This paper asks what might happen when two modern features of life insurance collide: policies with lapse-supported premiums, where future lapse surplus is committed in advance; and {adverse selection, which may reduce overall lapse rates. Our particular example is a ban on insurers' use of genetic test results.} 

% where limits may be placed on insurers' use of test results.

\subsection{Lapse-supported Premiums}
\label{sec:Outline0}

Booking future profits, before they have actually been earned, has been the end of many a financial institution. If a life insurer, upon selling a thirty-year policy (say), calculated its expected profit and distributed it to shareholders there and then, it would be asking for trouble. The avoidance of such risk, arguably, became the {\em raison d'etre} of the actuarial profession, {when} William Morgan faced down the proprietors of the young Equitable Life \citep{ogborn1962}. 

Distributing profits in cash or bonus form does at least require the profits to be declared as such, so the impact on the balance sheet can be {seen}. More subtle and less visible is distributing unearned {and uncertain} future profits to the very policyholders whose policies are supposed to earn them, by reducing contractual premiums. Such is the case with {\em lapse-supported premiums}. 

{Surrender values may be set at a low level},  to discourage lapsing and recover expenses. At short durations expenses are unlikely to have been fully recovered, but at longer durations the policy value or asset share may exceed the surrender value, and positive {lapse surplus} is earned. This is the future profit that may be applied to reduce the contractual premiums of lapse-supported policies. 

\begin{bajlist}

\item With-profit policies may benefit from lapse surplus, but only {\em retrospectively}, once it has been earned. 

\item Lapse-supported policies, generally non-profit, benefit from lapse surplus {\em prospectively}, before it has been earned. 

\end{bajlist}

Simply paying low or nil surrender values, as is usual {under contracts with small reserves} such as term insurance, may generate lapse surplus, and therefore support the business, but on its own that is not lapse-support. To meet the definition, the surplus must be anticipated at outset in the premium basis, and applied to reduce the contractual premiums.

Assuming that lapses are also anticipated in the valuation basis for lapse-supported business (to do otherwise would be rather odd), lapse surplus takes on exactly the same form as mortality surplus. Having a negative sum at risk {or death strain at risk}, similar to an annuity policy, the insurer makes a loss if lapse rates are {less than anticipated} (see Section \ref{sec:OutlineI}). Anything that encourages `sticky' policyholders, with lower lapse rates, then becomes a source of risk. 

\markright{Lapse Supported Life Insurance and Adverse Selection}

This leads at once to an economic argument: lower premium rates should encourage lower lapse rates, so we need a {joint model of lapse rates and premium rates}. This is a very interesting but difficult problem, which we have to leave for future research. Nevertheless, the contracts we consider (`Term to 100' contracts, see Section \ref{sec:TermTo100}) actually exist, and so does the new factor that may influence lapse rates (genetic testing, see Sections \ref{sec:LapseRisk} and \ref{sec:IntroGeneticTesting}). Moreover, that factor itself may {trigger} exogenous forces (the secondary market in insurance contracts).  It is still useful to ask the more primitive but practical question: how sensitive is profitability to lapse rates different from those assumed in the premium and valuation bases?

In this Section \ref{sec:Intro}, we sketch an outline of the paper. Section \ref{sec:TermTo100} defines `Term to 100' contracts, used throughout as an example, {and Section \ref{sec:Notation} introduces the required notation}. Section \ref{sec:OutlineI} defines lapse surplus, including notation, and Section \ref{sec:OutlineII} considers ownership of the lapse surplus, and how it may be distributed. Section \ref{sec:Funding} states our main mathematical result on the funding of lapse-support, which we believe to be new. 
Section \ref{sec:LapseRisk} introduces adverse selection, in its actuarial sense, and Section \ref{sec:IntroGeneticTesting} introduces genetic testing as a fruitful adverse selection risk, if restrictions are placed on insurers' use of genetic test results. This may lead to the creation of a new and `sticky' subpopulation with high mortality and low lapse rates. Section \ref{sec:IntroPlan} sets out the plan of the paper.

%-------------------------------------------------------------

\subsection{`Term to 100' Contracts}
\label{sec:TermTo100}

The canonical example of a lapse-supported contract is the Canadian `Term to 100' contract. In effect a whole-life assurance, technically it is a non-profit endowment contract maturing at age 100. Its main feature is the absence of cash or surrender values on lapse. However, lapse rates are included in the premium basis, so premiums are reduced, often very substantially so; see the comments in Section \ref{sec:Literature} and the example in Section  \ref{sec:Examples}. Variants of this contract are also common in the USA, the details depending on statewise regulations. 

%-------------------------------------------------------------

\subsection{{Notation}}
\label{sec:Notation}

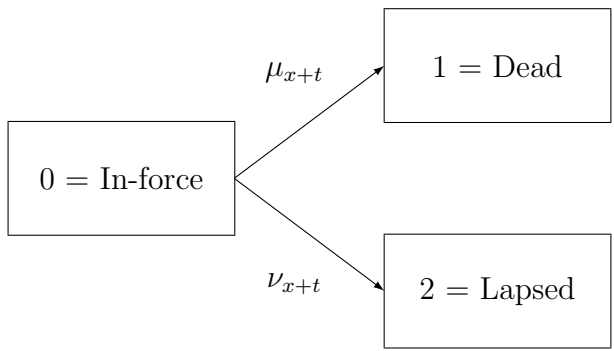
\begin{figure}
\begin{center}
\begin{picture}(95,50)
\put(50,30){\framebox(30,15){}}
\put(50,0){\framebox(30,15){}}
\put(0,15){\framebox(30,15){}}
\put(30,22.5){\vector(4,-3){20}}
\put(30,22.5){\vector(4,3){20}}
\put(15,22.5){\makebox(0,0)[c]{0 = In-force}}
% \put(15,20){\makebox(0,0)[c]{Prob = $p < 1$}}
\put(65,7.5){\makebox(0,0)[c]{2 = Lapsed}}
% \put(65,5){\makebox(0,0)[c]{Prob = 0}}
\put(65,37.5){\makebox(0,0)[c]{1 = Dead}}
% \put(65,35){\makebox(0,0)[c]{Prob = 1}}
\put(38,36.5){\makebox(0,0)[c]{$\mu_{x+t}$}}
\put(38,8.5){\makebox(0,0)[c]{$\nu_{x+t}$}}
\end{picture}
\end{center}
\caption{\label{fig:Lapses} A model of transfers between states representing in-force life insurance, death and lapsation.}
\end{figure}

We define below notation used throughout the paper, with the convention that dashed quantities (e.g. $\mu'_{x+t}$) represent the experience basis. {We use the continuous-time Markov model illustrated in Figure \ref{fig:Lapses}, with transition intensities as shown, and} assume a continuous-time model of cashflows, in which payments to the insurer are positive. Expenses are excluded for simplicity.

\begin{center}
\begin{tabular}{ll}
$n$ & Policy term (possibly $\infty$) \\
$x$ & The age at inception of a life insurance policy \\
$t$ & Policy duration \\
$\delta_t$ & {Force of interest} on premium or valuation basis \\
{$\delta'_t$} & {{Force of interest} on experience basis} \\
$\varphi(t)$ & Discount factor allowing for survivorship \\
$\mu_{x+t}$ & Mortality hazard rate on premium or valuation basis \\
$\mu'_{x+t}$ & Mortality hazard rate on experience basis \\
$\nu_{x+t}$ & Lapse hazard rate on premium or valuation basis \\
$\nu'_{x+t}$ & Lapse hazard rate on experience basis \\
$P(t)$ & Premium rate {per year} payable at duration $t$, without lapse-support \\
$P^*(t)$ & Premium rate {per year} payable at duration $t$, with lapse-support \\
$S(t)$ & Sum insured payable on death at duration $t$ \\
$C(t)$ & Surrender value payable on lapse at duration $t$ \\
$M$    & Maturity value payable at the end of the term (possibly zero) \\
$V(t)$ & Policy value at duration $t$, without lapse-support \\
$V^*(t)$ & Policy value at duration $t$, with lapse-support \\
$W(t)$ & Rate of surplus emerging at duration $t$, without lapse-support \\
$W^*(t)$ & Rate of surplus emerging at duration $t$, with lapse-support.
\end{tabular}
\end{center}

%-------------------------------------------------------------
%-------------------------------------------------------------

%-------------------------------------------------------------
%-------------------------------------------------------------

\subsection{Lapse {and Mortality} Surplus}
\label{sec:OutlineI}

% Lapse surplus arises when a policyholder terminates a policy at before the end of the term, and receives a benefit, against which the insurer can offset the policy value being held. The difference, positive or negative, is the net cashflow per lapsing policy. Lapse surplus is then the actual (experienced) lapse cashflow minus that anticipated in the premium or valuation basis. 

\noindent Premium rates and policy values are found by solving Thiele's equation \citep{dickson2020, hoem1988} see Section \ref{sec:Thiele}. Excluding lapses, Thiele's equation is:

\begin{equation}
\frac{d}{dt} V(t) = \delta_t \, V(t) + P(t) - \mu_{x+t} \, (S(t) - V(t)) \label{eq:ThieleWithout}
\end{equation}

\noindent and we obtain $P(t)$ and $V(t)$ by solving it with boundary values $V(0)=0, V(n)=M$. We assume that the form of $P(t)$ is specified in the contract (for example, level premiums throughout the term) so that Thiele's equation has a unique solution with the stated boundary values.

As the experience develops and differs from the valuation basis in equation (\ref{eq:ThieleWithout}), surplus at rate $W(t)$ emerges as a balancing item, as follows:

\begin{equation}
\frac{dV(t)}{dt} + W(t) = \delta'_t \, V(t) + P(t) - \mu'_{x+t} \, (S(t) - V(t)) - \nu'_{x+t} \, (C(t) - V(t)). \label{eq:ThieleExperienceWithout}
\end{equation}

\noindent Hence, by subtracting (\ref{eq:ThieleWithout}) from (\ref{eq:ThieleExperienceWithout}):

\begin{equation}
W(t) = (\delta'_t - \delta_t) \, V(t) - (\mu'_{x+t} - \mu_{x+t}) \, (S(t) - V(t)) - \nu'_{x+t} \, (C(t) - V(t)). \label{eq:FullSurplusWithout}
\end{equation}

% \noindent (\ref{eq:FullSurplusWithout}) shows that mortality is not the only contingency affecting the financial outcome, and need not be the most important, whatever the reason for entering into the contract in the first place. 

Insurers will usually try to avoid losses on lapse, by keeping  lapse rates as low as possible, {or} ensuring that $C(t) \le V(t)$ where possible. However, lapse rates are typically much higher than mortality rates (except at very high ages) so equation (\ref{eq:FullSurplusWithout}) shows that lapse surplus can be substantial, perhaps much more significant than mortality surplus. From this point on, we assume that experienced interest and mortality are as in the valuation basis, so $\delta'_t = \delta_t$ and $\mu'_{x+t} = \mu_{x+t}$, and equation (\ref{eq:FullSurplusWithout}) simplifies to:

\begin{equation}
W(t) = - \nu'_{x+t} \, (C(t) - V(t)). \label{eq:LapseSurplusWithout}
\end{equation}

\noindent This will simply accrue to the insurer's estate, unless a different owner is identified, and action taken to distribute the surplus to that owner.

%-------------------------------------------------------------

\subsection{Ownership and Distribution of Lapse Surplus}
\label{sec:OutlineII}

If lapse surplus is owned by shareholders or with-profit policyholders, then it can be distributed {\em via} the bonus system in use. We call this the {\em retrospective} distribution of lapse surplus. With-profit policyholders who have generated the surplus will benefit from it as a class, although there may be redistribution within the class. This is beyond the scope of this paper.

In the case of non-profit policyholders, retrospective methods of distribution are not available. {\em Prospective} distribution of lapse surplus means anticipating it in the pricing basis, and reducing the contractual premium, for {the same} benefits. {That is, we start with a non-profit contract with premium rate $P(t)$ at time $t$, assuming no future lapses, which will earn lapse surplus at rate $W(t)$ at duration $t$ (equation (\ref{eq:FullSurplusWithout}) or (\ref{eq:LapseSurplusWithout})). The {anticipated} lapse surplus {may} be respread over the premium-paying term} to pay for a reduction of the premium rate to $P^*(t) < P(t)$, with a consequent change of policy values, $V^*(t) \not= V(t)$. These quantities may be found by solving Thiele's equation allowing for lapses:

\begin{equation}
\frac{d}{dt} V^*(t) = \delta_t \, V^*(t) + P^*(t) - \mu_{x+t} \, (S(t) - V^*(t)) - \nu_{x+t} \, (C(t) - V^*(t)) \label{eq:ThieleWith}
\end{equation}

\noindent with boundary values $V^*(0)=0, V^*(n)=M$, and {suitable constraints on the form of $P^*(t)$} {(compare with equation (\ref{eq:ThieleWithout}))}

\begin{bajlist}

\item The form of equations (\ref{eq:ThieleWithout}) and (\ref{eq:ThieleWith}) does not, by itself, guarantee that $P^*(t) < P(t)$, but this will usually be the case for a conventional contract {with premium rate $P^*(t)$} where $0 \le C(t) \le V^*(t)$.

\item In cases where $P^*(t) < P(t)$ we say that the contract is {\em lapse-supported}.

\item {Lapse-supported business will make a loss if {experienced} lapse rates are below those assumed in the premium basis.}

\end{bajlist}

In much the same way as competition in the 19th century drove select mortality from being a source of surplus to an element of the premium basis, so competition in the 20th century turned lapsation into an element of the premium basis. {It is routine to include lapsing in profit-testing and sensitivity analysis, and for an element of lapse-support to be present in realized premiums. Our interest lies in a more extreme set of circumstances.

\begin{bajlist}

\item Policy values $V^*(t)$ are large, but there are no surrender values, $C(t)=0$. Our main example is the `Term to 100' contract, see Section \ref{sec:TermTo100}.

\item There are restrictions on medical underwriting that expose the insurer to adverse selection. Our example is genetic testing, see Sections \ref{sec:LapseRisk} and \ref{sec:IntroGeneticTesting}.

\item Adverse selection may be {boosted} by exogenous forces, encouraging exceptionally large sums insured to be taken out (`speculative adverse selection', see Section \ref{sec:LapseRisk}) and low rates of lapsing. Our example is the possibility of the life settlement industry encouraging the creation of such policies as investments.

\end{bajlist}} % END RED

We include consideration of a third strategy for dealing with lapse surplus, which is to avoid it completely. {There are many ways to do this; we choose a form of contract such} that $V(t) = C(t) = 0$ at all durations. This is achieved by having a premium rate equal to $\mu_{x+t} \, S(t)$ at duration $t$, and setting $C(t)=0$ at all durations. Then from equation (\ref{eq:FullSurplusWithout}), lapse surplus is zero whatever the lapse rates may be. We do not know of any free-standing contracts written on this basis but it closely resembles a common method of levying charges for life cover under unit-linked contracts. We call this form of contract a `current-cost policy'.

\newpage

%-------------------------------------------------------------

\subsection{Funding Lapse-Supported Premiums}
\label{sec:Funding}

{Define:

\begin{equation}
\varphi(t) = \exp \left( -\int_0^t (\delta'_r+\mu'_{x+r}+\nu'_{x+r}) \, dr \right) \label{eq:DiscountFactor}
\end{equation}

\noindent to be the discount factor at duration $t$, allowing for survivorship on the experience basis.} Our main result is an expression for the cost of lapse-supported premiums:

\begin{proposition} \label{prop:PropMain}
With the notation above, the expected present value {(EPV)} of the premium reduction is:

\begin{equation}
\int_0^{n} \varphi(t) \, (P(t)-P^*(t)) \, dt = \int_0^{n} \varphi(t) \, \nu'_{x+t} \, (V(t) - V^*(t)) \, dt + \int_0^{n} \varphi(t) \, \nu_{x+t} \, (V^*(t) - C(t)) \, dt. \label{eq:PremiumReduction}
\end{equation}

\end{proposition}

\noindent {This splits the cost of the premium reduction, on the left-hand side, into contributions from the expected cash surplus on lapse (second term on the right-hand side) and the realized difference in policy values released on lapse (first term of the right-hand side). We have not found this result in the literature.}

This proposition has two simple corollaries of general interest.

\begin{corollary}

In the case that $\nu'_{x+t} = \nu_{x+t}$, {so that experienced lapses are as assumed in the premium basis}, equation (\ref{eq:PremiumReduction}) simplifies as follows:

\begin{equation}
\int_0^{n} \varphi(t) \, (P(t)-P^*(t)) \, dt = \int_0^{n} \varphi(t) \, \nu_{x+t} \, ( V(t) - C(t) )\, dt \label{eq:PremiumReductionSimplified}
\end{equation}
 
\noindent so the cost of lapse-support is the EPV of the net cashflows of lapsed contracts. 

\end{corollary}

\begin{corollary} \label{cor:FY}

Suppose the contract has non-lapse-supported premium rate $P(t)$. Let $W(t)$ be the rate at which surplus emerges given policy values $V(t)$, lapse rate $\nu'_{x+t}$ and premiums $P(t)$. Now suppose the valuation basis is taken to be a net premium valuation\footnote{\label{footnote:NP} {Under a net premium valuation basis, policy values take the form: $$\mbox{EPV[Future Benefits]} - \mbox{EPV[Future {\em Net} Premiums]}$$ where the net premiums are calculated on the valuation basis, which need not be the same as the premium basis. See \cite{fisher1965} for example.}} with net premium rate $P^*(t)$ and policy values $V^*(t)$, and let $W^*(t)$ be the rate at which surplus emerges. Then:

\begin{equation}
\int_0^{n} \varphi(t) \, W(t) \, dt = \int_0^{n} \varphi(t) \, W^*(t) \, dt 
\end{equation}

\noindent so the EPV of the emerging surplus is the same under either valuation basis.

\end{corollary}

{Corollary \ref{cor:FY} is a special case of the result that the EPV on the experience basis of the total emerging surplus does not depend on the valuation basis. This result is well-known in jurisdictions where the valuation basis is not constrained to be the same as the premium basis (see \cite{fisher1965} for a typical statement). It appears to be less well-known elsewhere, and therefore missing from the modern literature on surplus. \cite{hacariz2024} provide a proof}. In fact this result would offer an alternative proof of Proposition \ref{prop:PropMain}.

Proofs of these results are in Section \ref{sec:Premiums}.

%-------------------------------------------------------------

\subsection{{Adverse Selection}}
\label{sec:LapseRisk}

Adverse selection may arise when there is an {\em information asymmetry} which the individual may exploit, as follows. The insurer charges a premium rate conditioned on the information known to them, call this $P^h(t)$. However, the individual has additional adverse information, unknown to the insurer, such that the appropriate premium rate is $P^a(t) > P^h(t)$. The difference in information is the individual's advantage. She may exploit it by obtaining insurance worth $\Mean[ \int \varphi(t) \, P^a(t) \, dt]$ at the lower price $\Mean[\int \varphi(t) \, P^h(t) \, dt]$. This is a form of adverse selection, against the insurer.

\cite{hacariz2020b} distinguished between two kinds of adverse selection:

\begin{bajlist}
\item {\em Precautionary adverse selection}: a higher probability of purchasing life insurance to meet normal needs.
\item {\em Speculative adverse selection}: taking out abnormally high sums insured, as a financial gamble exploiting the information advantage.
\end{bajlist}

\noindent As an example of (b), \cite{howard2014} assumed that 75\% of individuals with an adverse genetic test result would purchase ten times the normal sum insured, see Section \ref{sec:IntroGeneticTesting}.

Additionally, if the premium basis is lapse-supported then the insurer risks loss if experienced lapse rates are less than {the rates $\nu_{x+t}$ assumed in the premium basis}. It follows that the insurer has no incentive to keep lapse rates below that level, indeed quite the opposite. Anything which would encourage policyholders to keep their policies in force is a threat to the business --- a form of adverse selection after the point of sale. Since high lapse rates may be a sign of poor selling practices, this creates a conflict of interest for the insurer, which has led to some controversy around `Term to 100' contracts, see Section \ref{sec:Conflict}. 

% Some aspects of adverse selection rightly fall under the purview of economics. However, this is {\em not} an economics paper. We do not try to explain lapse rates by means of equilibrium or other econometric models. We have no data that would allow such models to be tested. It is still useful to take certain patterns of lapse behavior as given, leading to adverse selection, and consider the consequences for insurance losses. In Section \ref{sec:NumExAdvSel} we compare losses under a model `Term to 100' contract, with and without lapse-support, under several adverse selection scenarios. We also compare these outcomes with those under a hypothetical `current-cost' contract (see Section \ref{sec:OutlineII}) which eliminates policy values and cashflow losses on lapse.

%-------------------------------------------------------------

\subsection{Genetic Testing}
\label{sec:IntroGeneticTesting}

An example of inadequate medical underwriting being forced upon insurers is a ban on the use of genetic test results in underwriting. Since the 1990s, insurers have warned of the potential impact on the industry of such a ban, see \cite{pokorski1995} for example. Most attention has been paid to {\em single-gene disorders}, in which (simplifying greatly) heritable variants of a particular gene lead to one or more well-defined diseases, which therefore appear to `run in families'.

\begin{bajlist}

\item By `genetic test' we mean the direct examination of DNA which identifies high-risk variants of a particular gene, an outcome called an `adverse' genetic test result.

\item The fact that gene variants are heritable means that an individual's chance of carrying a risky variant may be deduced from the pattern of disease in blood relatives, called a `family medical history'. Therefore, as long as insurers may use this information, they are not completely ignorant of the risk, even if they may not use genetic test results. Genetic test results are relatively new, but family medical history has been used by insurers for a very long time.

\end{bajlist}

The example is real; the political and legislative background to such moves has been discussed at length elsewhere, see \cite{prince2019} or \cite{golinghorst2022}. Moreover, there have been attempts to gauge the impact of genetic testing on the life insurance industry \citep{macdonald2011, howard2014, lombardo2018}, some of which involve lapse-supported contracts. These are described in Appendix 2.

%-------------------------------------------------------------

\subsection{{Aim and} Plan of this Paper}
\label{sec:IntroPlan}

We try to answer two main questions: (a) how do lapse-supported premiums affect an insurer's exposure to the risk of adverse selection?; and (b) how robust are the conclusions of models used recently to illustrate the impact of banning insurers' access to genetic test results? In doing so we identify three ways of dealing with lapse surplus: (a) eliminating it by design; (b) disposing of it retrospectively; and (c) disposing of it prospectively (through lapse-supported premiums). 

% {This is {\em not} an economics paper. We do not try to explain lapse rates by means of equilibrium or other econometric models. We have no data that would allow such models to be tested. It is still useful to take certain patterns of lapse behavior as given, leading to adverse selection, and consider the consequences for insurance losses.}

In Section \ref{sec:LapseSupport} we briefly review the literature relevant to lapse-supported premiums, adverse selection and the secondary market, focusing on actuarial questions, and leaving aside economic questions such as efficiency and equilibrium. We give a numerical example of a `Term to 100' contract in Section \ref{sec:Examples}. In Section \ref{sec:Analysis} we analyze life insurance pricing and reserving, {with and without lapse-supported premiums}. {In Section \ref{sec:ActModelLSP} we analyze rates of adverse selection loss into its lapse and mortality components, and show how these  are changed by lapsing behavior. We give numerical examples in Section \ref{sec:NumExAdvSel}, with some reference to genetic testing, and our conclusions are in Section \ref{sec:Conclusions}. {Some proofs, and some background on genetic testing, which {motivates} this study, are given in the Appendices}. 

}

%-------------------------------------------------------------
%-------------------------------------------------------------
%-------------------------------------------------------------

\section{Lapse-supported Premiums}
\label{sec:LapseSupport}

\subsection{Literature on Lapse-supported Premiums}
\label{sec:Literature}

There is a modest literature on lapsation. \cite{shamsuddin2022} is a good source of references. Broadly, papers fall into four groups.

\begin{bajlist}

\item Studies reporting estimates of lapse rates.

\item Studies modeling lapse rates as functions of covariates, including measures of confidence and significance.

\item Studies testing hypotheses about what drives lapses, chiefly the Emergency Fund Hypothesis, the Interest Rate Hypothesis, and the Policy Replacement Hypothesis.

\item Other studies, including a small number concerned with lapse-supported premiums, and others concerned with the secondary market which mention the influence of lapse-supported premiums. The literature is sparse and not much is peer-reviewed. As \cite{gottlieb2021} say, ``Making a profit from policies that lapse is a taboo topic in the life insurance industry'' and ``$\ldots$ insurers are naturally tight-lipped about their pricing strategies.''.

\end{bajlist}

\noindent We will confine our attention to the relevant papers in (d). See \cite{eling2013} and \cite{eling2014} and references therein for publications in (a), (b) and (c).

The risks of lapse-supported premiums have been discussed anecdotally.  For example, \cite{gleen2005} and \cite{gottlieb2021} cite  Mr. M. Mahony at a Society of Actuaries meeting in October 1998 describing a 30-year term insurance policy making a profit (expected present value) of \$103,000 if the expected lapse assumptions were met, but a loss of \$942,000 if there were no lapses. Such a large swing may look surprising, but is consistent with the standard deviations of adverse selection costs in the examples in Section \ref{sec:ActModelLSP}. \cite{gleen2005} also noted that lapse-supported pricing was a significant reason for large differences between prices quoted by different insurers for (apparently) the same benefits. 

\cite{record1987}, a panel discussion on reserving for lapse-supported business, has many valuable insights. {Otherwise, there appears to be no published study of the actuarial aspects of lapse-support. \cite{gottlieb2021} could cite primarily anecdotal evidence, and their own calculations based on surveyed premium rates that suggested heavy reliance on lapse-support in the USA.} \cite{gatzert2009} was the first quantitative study to assume  that lapse behavior will be affected by the secondary market.  They concluded that if policyholders whose health had worsened during the policy term lapsed at lower rates than normal, insurers would see significantly reduced surrender profits, or even losses. Premiums {in their model} were not lapse-supported, and surrender values were paid upon lapse, but they concluded that:

\begin{quote} 
\small 
``In the long run, both consumers and life insurance carriers will benefit from a competitive secondary market. $\ldots$ However, life insurers will need to abandon lapse-supported pricing, which could also aid in reducing the volatility of their profits'' 
\end{quote} 

\noindent (in which respect see Table \ref{table:SecondMoments1}).

%-------------------------------------------------------------

\subsection{`Term to 100' Contracts and the Insurer's Conflict of Interest}
\label{sec:Conflict}

The conflict of interest posed by planning for lapse surpluses was neatly summed up in this exchange from \cite{record1987}, a panel discussion:

\begin{quote}
\small

\noindent ``{\sc MR MCFARLANE:} $\ldots$ What are the ethics of designing a product to encourage lapses?

\smallskip

\noindent {\sc MR GOLD:} This is a problem that seems to worry people in the U.S. I see no problem with this product at all, as long as everything is clearly explained on day one. This is what you get, you get no cash value or paid-up value. $\ldots$''
\end{quote}

\noindent \cite{gleen2005} mentioned that: ``Advocates of lapse-supported pricing can find their best case in Canada, where valuation and nonforfeiture laws allow product designs that are not possible in the U.S.''. This was an oblique reference to the Canadian `Term to 100' contract. Also in \cite{record1987}, Mr R. W. MacDonald said:

\begin{quote}
\small
 ``Then of course, there was perhaps our most famous, or infamous --- depending on your point of view --- the `Term to 100' product. The generic type required level premiums to age 100 and provided no value whatsoever upon failure to pay premiums.'' 
\end{quote}

The argument in support of lapse-support is that policyholders can keep cover in force while it is needed, for a term not known in advance, and then drop it when it is no longer needed, more cheaply than if premiums were not lapse-supported.

This conflict of interest is sometimes assumed to underlie a strong antipathy on the part of insurers towards the secondary market. By offering policyholders an alternative to lapsing, it is supposed, life settlement companies disrupt the underlying basis of lapse-supported pricing. Some authors go so far as to view the secondary market as a welcome prophylactic against lapse-supported premiums, for example \cite{doherty2003} suggest it will ``$\ldots$ keep incumbent insurers from  the unfair, and ultimately unworkable, practice of using high lapse expectations to underprice certain policies.''

%-------------------------------------------------------------

\subsection{An Example of Lapse-supported Premiums}
\label{sec:Examples}

Table \ref{table:lsp1} demonstrates the main features of lapse-supported premiums using as a simple example, a non-profit endowment contract written for a male age 35 and maturing at age 100. If no surrender values are paid, this is a typical `Term to 100' benefit.
 
Premiums are payable continuously, death benefits are payable immediately on death, the sum insured is \pounds 250,000, and the {first-order technical basis for} pricing  is as follows: force of interest 0.03, 0.06 or 0.09 {\em per annum} as in Table \ref{table:lsp1}; mortality GM82 Males (a Danish standard table); no expenses. {Premiums are calculated allowing for no lapses (no lapse-support) or lapse rates of 0.03 or 0.06 {\em per annum}.} Policy values are calculated on the same basis {as premiums}.

\begin{table}
\begin{center}
\caption{\label{table:lsp1} Examples of lapse-supported premiums for 65-year endowment contracts (`Term to 100'), males age 35, sum insured \pounds 250,000, annual premiums. Profit is \% of {EPV[Premiums without lapse-support] with lapses occurring. Loss is \% of EPV[Premiums with lapse-support] with no lapses occurring}.}
\small
\begin{tabular}{lllllrr} \\
SV as \% &          &         & {Premium}   & {Premium}     &        & \\
Policy   & Force of & Lapse   & {No Lapse-} & {With Lapse-} & Max.     & Max. \\ 
Value    & Interest & Rate    & {Support}   & {Support}     & Profit & Loss \\[0.5ex]
(\%) & & & & & (\%) & (\%) \\[0.5ex]
  50 &   0.03 &   0.03 &    3744.44 &    2919.43 &      38.00 & $-$56.52 \\
  50 &   0.03 &   0.06 &    3744.44 &    2321.62 &      57.64 & $-$122.57 \\[0.5ex]
  50 &   0.06 &   0.03 &    2321.62 &    1892.86 &      31.68 & $-$45.30 \\
  50 &   0.06 &   0.06 &    2321.62 &    1586.02 &      48.09 & $-$92.76 \\[0.5ex]
  50 &   0.09 &   0.03 &    1586.02 &    1365.40 &      24.01 & $-$32.32 \\
  50 &   0.09 &   0.06 &    1586.02 &    1205.15 &      37.03 & $-$63.21 \\[0.5ex]
   0 &   0.03 &   0.03 &    3744.44 &    2321.62 &      38.00 & $-$61.29 \\
   0 &   0.03 &   0.06 &    3744.44 &    1586.02 &      57.64 & $-$136.09 \\[0.5ex]
   0 &   0.06 &   0.03 &    2321.62 &    1586.02 &      31.68 & $-$46.38 \\
   0 &   0.06 &   0.06 &    2321.62 &    1205.15 &      48.09 & $-$92.64 \\[0.5ex]
   0 &   0.09 &   0.03 &    1586.02 &    1205.15 &      24.01 & $-$31.60 \\
   0 &   0.09 &   0.06 &    1586.02 &     998.73 &      37.03 & $-$58.80 
\end{tabular}
\end{center}
\end{table}

Note the following features in Table \ref{table:lsp1}.

\begin{bajlist}

% \item The reduction in premium can be very large,  because of the long term and high lapse rates.

\item When the surrender value is 100\% of the policy value, the premiums with and without lapse-support are equal, and the `Profit' and `Loss' columns are zero, so we omit these cases from the table.

\item When less than 100\% of the policy value is paid as the surrender value, the lapse-supported premium can be considerably reduced, giving a competitive advantage. This is because of the long policy term and high lapse rates. % This is an expression of  Lidstone's theorem, allowing for lapses. 

\item Lapse-supported premium rates are equal for certain combinations of interest and lapse rates. This is a consequence of equation (\ref{eq:Thiele3}).

\item If the office charges the full premium (not lapse-supported) but lapses are experienced, it may generate surplus, depending on the surrender values paid. The maximum possible surplus, when surrender values are zero, is shown as a percentage of the expected present value (EPV) of premiums in the column headed `Max. Profit'.

\item If the office charges the lapse-supported premium but experiences no lapses, it will make a maximal loss, shown as a percentage of the EPV of premiums in the column headed `Max. Loss'.

\end{bajlist}

Note that, because of (c) above, we cannot tell whether any given premium rate is lapse-supported, or to what degree, without knowing the premium basis.

%------------------------------------------------------------
%------------------------------------------------------------
%------------------------------------------------------------

\section{Pricing and Reserving with Allowance for Lapses}
\label{sec:Analysis}

\subsection{Thiele's Equation in Pricing and Reserving}
\label{sec:Thiele}

In equation (\ref{eq:ThieleWithout}) we stated Thiele's equation for a contract without lapse-support, that is, assuming $\nu_{x+t}=0$. 

\begin{bajlist}

\item The terminal boundary condition is $V(n) = M$, where $M=0$ defines a pure protection policy with no maturity value, and $M>0$ defines a `permanent' assurance with maturity value $M$.

\item For pricing under the equivalence principle, {we find $P(t)$ such that, in addition, the initial boundary condition $V(0)=0$ is satisfied.} In general there will be infinitely many such premium functions, and our choice will be determined by the policy design. Common premium functions are single premium, level premiums, arithmetically or geometrically increasing premiums, and premiums for a limited term shorter than the policy term. 

\item For calculating policy values with a given premium function $P(t)$ we solve equation (\ref{eq:ThieleWithout}) backwards from the terminal condition $V(n)=M$. 

\end{bajlist}

%-------------------------------------------------------------

\subsection{Allowing for Lapses in Pricing and Reserving}
\label{sec:Lapses}

To incorporate lapses we define the three-state model shown in Figure \ref{fig:Lapses}, with lapse hazard rate (intensity)  $\nu_{x+t}$ at time $t$. On lapsing at time $t$ the surrender value is $C(t)$ (possibly zero). Then Thiele's equation for pricing and reserving was given in equation (\ref{eq:ThieleWith}).  We confine attention to policy designs where $0 \le C(t) \le V^*(t)$ at all times $t \ge 0$. 

A special case, noted by \cite{lidstone1905} is that $C(t)$ is defined to be a proportion $k$ of the policy value $V^*(t)$ $(0 \le k \le 1)$. Then from equation (\ref{eq:ThieleWith}):

\begin{eqnarray}
\frac{d}{dt} V^*(t) & = & \big( \delta_t + (1-k) \, \nu_{x+t} \big) \, V^*(t) + P^*(t) - \mu_{x+t} (S(t) - V^*(t)). \label{eq:Thiele3}
\end{eqnarray}

\noindent  Equation (\ref{eq:Thiele3}) is functionally identical to equation (\ref{eq:ThieleWithout}), with the force of interest increased by $(1-k) \, \nu_{x+t}$. For example, if lapse rates are a constant 0.03 per year and surrender values are 50\% of policy values, then effectively the force of interest in equation (\ref{eq:ThieleWithout}) is increased by 0.015 per year. This result provides a useful rule of thumb, but does not imply that the management of lapse risk and asset return risk are equivalent, which clearly they are not.

%------------------------------------------------------------

\setcounter{proposition}{0}
\setcounter{corollary}{0}

\subsection{Lapse-supported Premiums}
\label{sec:Premiums}

In this section we show how the premium reductions are funded by anticipating lapse surpluses. Our main result is the following:

\begin{proposition}
The expected present value, on the experience basis, of the premium reduction due to including lapse rates $\nu_{x+t}$ in the premium basis, is:

\begin{equation}
\int_0^{n} \varphi(t) \, (P(t)-P^*(t)) \, dt = \int_0^{n} \varphi(t) \, \nu'_{x+t} \, (V(t) - V^*(t)) \, dt + \int_0^{n} \varphi(t) \, \nu_{x+t} \, (V^*(t) - C(t)) \, dt. \label{eq:PremiumReduction2}
\end{equation}

\end{proposition}

\noindent {\em Proof}: Following \cite{linnemann1993} (for example), consider the derivative of $\varphi(t) \, V(t)$:

\begin{eqnarray}
\frac{d}{dt} \, \big( \varphi(t) \, V(t) \big) & = & - \varphi(t) \, ( \delta'_t + \mu'_{x+t} + \nu'_{x+t} ) \, V(t) + \varphi(t) \, \frac{d V(t)}{dt} \nonumber \\
& = & - \varphi(t) \, ( \delta'_t + \mu'_{x+t} + \nu'_{x+t} ) \, V(t) + \varphi(t) \, \big( \delta_t \, V(t) + P(t) - \mu_{x+t} \, (S(t) - V(t)) \big) \nonumber \\
& & \label{eq:DerivWithout}
\end{eqnarray}

\noindent upon inserting equation (\ref{eq:ThieleWithout}). Rearranging this, we get:

\begin{equation}
\frac{d}{dt} \, \big( \varphi(t) \, V(t) \big) = - \varphi(t) \, V(t) \, \big( (\delta'_t - \delta_t) + (\mu'_{x+t} - \mu_{x+t}) + \nu'_{x+t}  \big) + \varphi(t) \, (P(t) - \mu_{x+t} \, S(t)). \label{eq:DerivWithout2}
\end{equation}

\noindent We shall ignore interest and mortality surplus, by assuming $\delta'_t=\delta_t$ and $\mu'_{x+t}=\mu_{x+t}$, so equation (\ref{eq:DerivWithout2}) simplifies to:

\begin{equation}
\frac{d}{dt} \, \big( \varphi(t) \, V(t) \big) = - \varphi(t) \, V(t) \, \nu'_{x+t} + \varphi(t) \, (P(t) - \mu_{x+t} \, S(t)). \label{eq:DerivWithout2b}
\end{equation}

\noindent Following the same steps for the derivative of $\varphi(t) \, V^*(t)$ we get:

\begin{eqnarray}
\frac{d}{dt} \, \big( \varphi(t) \, V^*(t) \big) & = & - \varphi(t) \, V^*(t) \, \big( (\delta'_t - \delta_t) + (\mu'_{x+t} - \mu_{x+t}) + (\nu'_{x+t} - \nu_{x+t}) \big) \nonumber \\
& & + \varphi(t) \, \big( P^*(t) - \mu_{x+t} \, S(t) - \nu_{x+t} \, C(t) \big). \label{eq:DerivWithout4}
\end{eqnarray}

\noindent Ignoring interest and mortality surplus, equation (\ref{eq:DerivWithout4}) simplifies to:

\begin{equation}
\frac{d}{dt} \, \big( \varphi(t) \, V^*(t) \big) = - \varphi(t) \, (\nu'_{x+t} - \nu_{x+t}) \, V^*(t) + \varphi(t) \, \big( P^*(t) - \mu_{x+t} \, S(t) - \nu_{x+t} \, C(t) \big). \label{eq:DerivWithout3}
\end{equation}

\noindent Note that $V(0)=V^*(0)=0$ and $V(n) = V^*(n) = M$, so that:

\begin{equation}
\int_0^n \frac{d}{dt} \, \big( \varphi(t) \, ( V^*(t) - V(t) ) \big) \, dt = 0. \label{eq:Zero}
\end{equation}

\noindent {Subtract equation (\ref{eq:DerivWithout2b}) from equation (\ref{eq:DerivWithout3})}, integrate over the policy term, apply equation (\ref{eq:Zero}), and on rearranging we get:

\begin{equation}
\int_0^n \varphi(t) \, (P(t) - P^*(t)) \, dt = \int_0^n \varphi(t) \, \nu'_{x+t} \, (V(t) - V^*(t)) \, dt + \int_0^n \varphi(t) \, \nu_{x+t} \, (V^*(t) - C(t)) \, dt. \label{eq:Loadings2}
\end{equation}

\hfill{$\Box$}

We state without {further} proof the following corollary:

\begin{corollary}

In the case that $\nu'_{x+t} = \nu_{x+t}$, equation (\ref{eq:Loadings2}) simplifies as follows:

\begin{equation}
\int_0^{n} \varphi(t) \, (P(t)-P^*(t)) \, dt = \int_0^{n} \varphi(t) \, \nu_{x+t} \, ( V(t) - C(t) ) \, dt.\label{eq:Loadings3}
\end{equation}
 
\end{corollary}

These results are fundamental to any analysis of lapse-supported premiums. 

\begin{bajlist}

\item Equation (\ref{eq:Loadings3}) shows that, if lapses are as expected ($\nu'_{x+t} = \nu_{x+t}$) then lapse-support is equivalent to capitalizing future lapse surplus, not in lump-sum form but respread over future premiums. The right-hand side of equation (\ref{eq:Loadings2}) shows the adjustments needed otherwise.

\item A contract with lapse-supported premiums therefore belongs to that class of financial contracts which bring forward future profits and recognize them in the balance sheet before they have been realized in revenue.

\item We assume in the above that that part of the policy value $V^*(t)$ of a lapsing policy which is {\em not} capitalized in the form of reduced premium, is paid as a surrender value to the policyholder. This is a convenient assumption but need not be the case; such surplus could be paid to proprietors or contribute to the estate, for example. All that is needed to accommodate this {explicitly} is more notation.

\item If interest and mortality surplus are present, equations (\ref{eq:Loadings2}) and (\ref{eq:Loadings3}) will have additional terms, not involving lapsing.

% \item We have defined two ways in which lapse surplus can be disposed of\footnote{There is an analogy with net {\em versus} gross premium valuations, {where} the difference (Gross Premium $-$ Net Premium) is the {\em premium loading}. The net premium valuation allows these to fall into surplus only as each future premium is paid, while the gross premium valuation capitalizes them into a lump sum, which is taken into surplus at outset.}.

% \begin{bajsublist}
% \item {\em Retrospectively}, allowing lapse surplus to emerge naturally as policies lapse, and then disposing of it as surrender values or otherwise.
% \item {\em Prospectively}, anticipating future lapses and `mortgaging' some or all of the surplus as premium reductions from outset.
% \end{bajsublist}

% \noindent At one extreme we have $C(t)=V^*(t)$ $(0 \le t \le n)$, which implies $P^*(t)=P(t)$ and $V^*(t)=V(t)$, and no lapse-support; and at the other we have all $C(t)=0$, mortgaging all {anticipated} lapse surplus.

\item For further developments of surplus emerging under the application of Thiele's equation see \cite{ramlau-hansen1991} and \cite{linnemann1993}.

\end{bajlist}

%-------------------------------------------------------------

\subsection{{Total Surplus and the Valuation Basis}}
% \subsection{Earning and Disposing of Lapse Surplus}
\label{sec:LapseSurplus}

Note that an analogue of equation (\ref{eq:Zero}) holds if any strictly positive discounting function $\tilde{\varphi}(t)>0$ is substituted for $\varphi(t)$. In particular, if we define:

\begin{equation}
\tilde{\varphi}(t) = \exp \left( - \int_0^t (\delta_r + \mu_{x+r} + \nu_{x+r} ) \, dr  \right)
\end{equation}

\noindent we also obtain equation (\ref{eq:Loadings3}). This might seem simpler, but discounting on the experience basis using $\varphi(t)$ leads to the following corollary of interest:

\begin{corollary}

Let the experienced lapse rate be $\nu'_{x+t}$. For brevity, we assume that $\delta'_t=\delta_t$ and $\mu'_{x+t}=\mu_{x+t}$, although the corollary is true without this assumption. Suppose that a contract has fixed surrender values $C(t)$. Without lapse support, it has {contractual} premium rate $P(t)$ and policy values $V(t)$, given by Thiele's equation (\ref{eq:ThieleWithout}). Let $W(t)$ be the rate at which surplus emerges. Now suppose the valuation basis is taken to be a net premium valuation {(see footnote \ref{footnote:NP})}, with net premium rate equal to the lapse-supported premium rate $P^*(t)$ and policy values $V^*(t)$, given by Thiele's equation (\ref{eq:ThieleWith}) with lapse rate $\nu_{x+t}$. Let $W^*(t)$ be the rate at which surplus now emerges. Then:

\begin{equation}
\int_0^{n} \varphi(t) \, W(t) \, dt = \int_0^{n} \varphi(t) \, W^*(t) \, dt \label{eq:EqualEPVSurpluses}
\end{equation}

\noindent so the EPV, on the experience basis, of the emerging surplus is the same under either valuation basis.

\end{corollary}

\noindent {\em Proof}: On the valuation basis equal to the premium basis, with policy values $V(t)$ and premium rate $P(t)$ we have:

\begin{equation}
\frac{dV(t)}{dt} + W(t) = \delta_t \, V(t) + P(t) - \mu_{x+t} \, (S(t) - V(t)) - \nu'_{x+t} \, (C(t)-V(t))
\end{equation}

\noindent and on subtracting equation (\ref{eq:ThieleWithout}) we get:

\begin{equation}
W(t) = - \nu'_{x+t} \, (C(t)-V(t)).
\end{equation}

\noindent On the alternative valuation basis with policy values $V^*(t)$ and {net} premium rate $P^*(t)$ we have:

\begin{equation}
\frac{dV^*(t)}{dt} + W^*(t) = \delta_t \, V^*(t) + P(t) - \mu_{x+t} \, (S(t) - V^*(t)) - \nu'_{x+t} \, (C(t)-V^*(t))
\end{equation}

\noindent and on subtracting equation (\ref{eq:ThieleWith}) we get:

\begin{equation}
W^*(t) = (P(t) - P^*(t)) - ( \nu'_{x+t} - \nu_{x+t}) \, (C(t)-V^*(t)).
\end{equation}

\noindent Hence:

\begin{eqnarray}
\int_0^n \varphi(t) \, (W^*(t)-W(t)) \, dt & = & \int_0^n \varphi(t) \, (P(t) - P^*(t)) \, dt - \int_0^n \varphi(t) \, ( \nu'_{x+t} - \nu_{x+t}) \, (C(t)-V^*(t)) \, dt \nonumber \\
& & \qquad + \int_0^n \varphi(t) \, \nu'_{x+t} \, (C(t)-V(t)) \, dt \\
& = & 0
\end{eqnarray}

\noindent on rearranging and using equation (\ref{eq:Loadings2}). \hfill{$\Box$}

In fact, Corollary 2 is a special case of a general result, well-known in literature of UK origin (see, for example, 
\cite[pp, 202--203]{fisher1965}, {and \cite{hacariz2024} for a recent proof)}. The result states that if we have two different valuation bases for a policy with fixed contractual payments, including the premiums, then the EPVs, calculated on the experience basis, of all the surpluses emerging under either basis, are equal\footnote{It is common practice in the UK to consider valuation bases other than the premium basis for life insurance contracts, resulting in different patterns of emerging surplus given a fixed  experience basis. This result states that the {timing of the emerging surplus  depends on the valuation basis, but its expected present value does not.} Almost all of the modern literature on surplus in life insurance (for example \cite{ramlau-hansen1988a, ramlau-hansen1991, linnemann1993, norberg1991}) assumes that the valuation and premium bases are the same (the `first-order technical basis').}. Note that we do not rely on this result in the above, instead proving it for our particular example.

The results of this section underlie our description, in Section \ref{sec:OutlineII}, of lapse-supported premiums distributing lapse surplus prospectively, rather than retrospectively through bonus or other means. In the next section we consider the third possible treatment of lapse surplus, elimination through policy design.

%-------------------------------------------------------------

\subsection{A Different Approach: Premiums Equal to Mortality Cost}
\label{sec:PremFunc}

% Level premiums give rise to policy values $V(t) > 0$ if mortality increases rapidly with age, as we always assume. Negative policy values are usually avoided in practice, and set to zero in the balance sheet, to avoid assets vanishing when policies lapse. Whether $V(t)$ is always increasing or sometimes decreasing depends on the death benefit function $S(t)$, maturity benefit $M$ and the mortality table. 

% We will have cause to consider, {in Section \ref{sec:Case3},} a premium function equal to the mortality cost at time $t$, namely {of the form}:

As an alternative to the usual assumption of level premium rates, we consider a premium function equal to the mortality cost at time $t$, namely {of the form}:

\begin{equation}
P(t) = \mu_{x+t} \, (S(t) - V(t)). \label{eq:MortCost}
\end{equation}

\noindent Then Thiele's equation, without lapses (equation (\ref{eq:ThieleWithout})) reduces to:

\begin{equation}
\frac{d}{dt} V(t) = \delta_t \, V(t) \label{eq:Thiele1b}
\end{equation}

\noindent or with lapses (equation (\ref{eq:ThieleWith}) and surrender values $C(t)=0$), to:

\begin{equation}
\frac{d}{dt} V^*(t) = (\delta_t + \nu_{x+t}) \, V^*(t) \label{eq:Thiele2b}
\end{equation}

\noindent which, with the initial conditions $V(0)=0$   $(V^*(0)=0)$, have the trivial solutions $V(t)=0$ $(V^*(t)=0)$ for all $t \ge 0$. Consequently we assume that surrender values are always zero with this premium function. By definition, these premiums are not lapse-supported.

This form of premium is not common for stand-alone protection policies, but is common, in a monthly discretized form, as an explicit mortality charge under unit-linked policies, where the unit fund value takes the place of $V(t)$ (or $V^*(t)$). Since such charges form part of the cashflows attributable to the insurer rather than the policyholder, it is proper to treat them as premiums in the calculation of surplus. 

For our purposes this premium function defines a third way to dispose of lapse cashflow surpluses, in addition to retrospectively and prospectively, namely to {\em eliminate them by design of the policy}.

\newpage

%-------------------------------------------------------------
%-------------------------------------------------------------
%-------------------------------------------------------------

\section{Analyzing the Costs of Adverse Selection}
\label{sec:ActModelLSP}

\subsection{Surplus in Inhomogeneous Populations}
\label{sec:ThieleInhom}

In this section we describe how adverse selection may arise, and give expressions for rates of mortality and lapse surplus arising in the three cases outlined below. Numerical illustrations will be given in Section \ref{sec:NumExAdvSel}.

We suppose the insured population consists of: (a) a large `normal' subpopulation, with the label $j=1$; and (b) a small `high-risk' subpopulation, with the label $j=2$, who are mistakenly charged the same ordinary premium as the larger group, leading to a loss. The loss may be exacerbated by the `high-risk' subpopulation exhibiting any of the following behaviors: {(a) being more likely to buy insurance; (b)} choosing higher sums insured than normal; and (c) being less likely to lapse policies\footnote{Appendix 2 outlines briefly how insurers may be at risk of adverse selection if they are denied access to genetic test results, and two approaches which have been taken to modeling the costs of such adverse selection, to which we may refer for motivation in this section.}.

We will consider a representative `Term to 100' contract, and analyze the EPV of losses in selected scenarios under the following three cases:

\begin{bajlist}
\item Case 1: level premiums, no lapse support;
\item Case 2: level premiums, with lapse support; and
\item Case 3: premiums equal to mortality cost;
\end{bajlist}

\noindent We assume that individuals in the `high-risk' subpopulation $j=2$ may exercise adverse selection in two ways. 

\begin{bajlist}
\item The proportion of high-risk individuals in the {\em insured} population at age $x$, denoted by $\pi_0$, may be much higher than the prevalence of high-risk individuals in the general population at age $x$. In conjunction with the normal sum insured $S$ this is `precautionary adverse selection' \citep[see also Section \ref{sec:IntroGeneticTesting}]{hacariz2020b}.
\item Each high-risk individual may have sum insured $\theta \times S$, where $\theta \ge 1$. If $\theta > 1$ this is `speculative adverse selection' \citep[see also Section \ref{sec:IntroGeneticTesting}]{hacariz2020b}.
\end{bajlist}

For brevity we ignore expenses, which are easily accommodated, and for simplicity we assume that $\delta_t = \delta > 0$, a constant. The following list summarizes the notation we use. 

\begin{center}
\begin{tabular}{ll}
$\delta$          & constant force of interest \\
$\pi_0$           & initial proportion of policies in the `high-risk' subpopulation \\
$\pi(t)$          & proportion of policies in `high-risk' subpopulation at time $t$ \\
$\mu_{x+t}$       & mortality hazard rate in valuation basis \\
$\nu_{x+t}$       & lapse hazard rate in valuation basis \\
$\mu_{x+t}^{(j)}$ & mortality hazard rate in subpopulation $j$ \\
$\nu_{x+t}^{(j)}$ & lapse hazard rate in subpopulation $j$ \\
${}_tp_x^{(j)}$   & survival probability from age $x$ in subpopulation $j$ \\
$W(t)$            & rate of surplus emerging per policy in force \\
$W^{(j)}(t)$      & rate of surplus emerging per policy in force in subpopulation $j$ \\
$\theta$          & multiple of `normal' sum insured purchased in subpopulation 2
\end{tabular}
\end{center}

The EPV of the total surplus at time $t$ in this model is:

\begin{equation}
\mbox{EPV[Total Surplus]} = (1-\pi_0) \int_0^t v^s \, {}_sp_x^{(1)} \,  W^{(1)}(s) \, ds + \pi_0 \int_0^t v^s \, {}_sp_x^{(2)} \, W^{(2)}(s) \, ds. \label{eq:TotSurp}
\end{equation}

\noindent From equation (\ref{eq:TotSurp}), adverse selection losses may be attributed to:

\begin{bajlist}
\item different survival probabilities {(see Section \ref{sec:Attrition})}; or
\item different rates of surplus emergence {(see Section \ref{sec:Uniform})}
\end{bajlist}

\noindent in the two subpopulations. % In this section we examine the rates at which adverse selection losses (surpluses) emerge in different circumstances.

%-------------------------------------------------------------

\subsection{Relative Attrition of the Subpopulations: {Survival Probabilities}}
\label{sec:Attrition}

Individuals in the `normal' and `high-risk' subpopulations at age $x$ are still policyholders at age $x+t$ with probabilities denoted by ${}_tp_x^{(1)}$ and ${}_tp_x^{(2)}$ defined as:

\[ {}_tp_x^{(1)} = \exp \left( - \int_0^t (\mu_{x+s}^{(1)} + \nu_{x+s}^{(1)} ) \, ds \right) \qquad \mbox{and} \qquad {}_tp_x^{(2)} = \exp \left( - \int_0^t (\mu_{x+s}^{(2)} + \nu_{x+s}^{(2)} ) \, ds \right) \]

\noindent respectively. Therefore the expected proportion in the `high-risk' subpopulation at age $x+t$, denoted by $\pi(t)$, is:

\begin{equation}
\pi(t) = \frac{\pi_0 \, {}_tp_x^{(2)}}{(1-\pi_0) \, {}_tp_x^{(1)} + \pi_0 \, {}_tp_x^{(2)}}. \label{eq:Attrition}
\end{equation}

%-------------------------------------------------------------

\subsection{Uniform and Differential Lapsing}
\label{sec:Uniform}

If lapse rates are the same (possibly zero) in the `normal' and `high-risk' subpopulations, we have {\em uniform lapsing}, or if they are different, we have {\em differential lapsing}. So, given  differential mortality, $\mu_{x+t}^{(2)} > \mu_{x+t}^{(1)}$, surplus arises in two stages, depending on lapsing behavior.

\begin{bajlist}
\item Stage 1: Under uniform lapsing, lapse rates vanish from equation (\ref{eq:Attrition}), and adverse selection occurs only because of the higher mortality in the `high-risk' subpopulation. We regard the resulting loss as `pure' mortality loss.
\item Stage 2: Under differential lapsing, $\nu_{x+t}^{(2)} < \nu_{x+t}^{(1)}$, and further losses arise which may truly be attributed to lapse behavior.
\end{bajlist} 

% Case 3, premium = mortality cost, is different in kind and is considered separately in Section \ref{sec:Case3}.

%-------------------------------------------------------------

\subsection{Rates of Surplus Under Adverse Selection}
\label{sec:SurplusAdverse}

Let $W(t)$ be the rate at which surplus is earned at time $t$ per policy in force. Appendix 1 shows details of calculating $W(t)$ in the three cases above. For convenience they are summarized in Table \ref{table:LossesTheory}, split into mortality and lapse surplus.

Table \ref{table:LossesTheory} sets out the rates of loss arising from mortality and from lapse behavior in the three cases, with uniform and differential lapsing. Case 1 is divided according to the payment of surrender values: no distribution, $C(t)=0$; and full distribution, $C(t)=V(t)$.

%-------------------------------------------------------------

\begin{landscape}
\begin{table}
\begin{center}
\caption{\label{table:LossesTheory} Examples of rates of adverse selection loss, attributable to mortality and lapses, per in-force policy. Sum insured in `high-risk' subpopulation $\theta \, S$. `Unif' means uniform lapsing $\nu_{x+t}^{(2)} = \nu_{x+t}^{(1)}$ and `Diff' means differential lapsing $\nu_{x+t}^{(2)} = 0$.}
\small
\begin{tabular}{llllll}
& & & & & \\
& & Surr & & Rate of Adverse Selection Loss & Rate of Adverse Selection Loss \\
Case   & Description & Value & Lapsing & Mortality & Lapses \\[0.5ex]
Case 1 & Level prem, no lapse support & 0      & Unif & $ - \pi(t) \, \theta \, ( \mu_{x+t}^{(2)} - \mu_{x+t} ) \, ( S - V(t) )$ & $ \big( (1-\pi(t)) \, \nu_{x+t}^{(1)} + \pi(t) \, \theta \, \nu_{x+t}^{(2)} \big) \, V(t)$ \\
Case 1 & Level prem, no lapse support & 0      & Diff & $ - \pi(t) \, \theta \, ( \mu_{x+t}^{(2)} - \mu_{x+t} ) \, ( S - V(t) )$ & $ (1-\pi(t)) \, \nu_{x+t}^{(1)} \, V(t)$ \\ 
Case 1 & Level prem, no lapse support & $V(t)$ & Unif & $ - \pi(t) \, \theta \, ( \mu_{x+t}^{(2)} - \mu_{x+t} ) \, ( S - V(t) )$ & nil \\
Case 1 & Level prem, no lapse support & $V(t)$ & Diff & $ - \pi(t) \, \theta \, ( \mu_{x+t}^{(2)} - \mu_{x+t} ) \, ( S - V(t) )$ & nil \\
Case 2 & Level prem, with lapse support & 0 & Unif & $ - \pi(t) \, \theta \, ( \mu_{x+t}^{(2)} - \mu_{x+t} ) \, ( S - V^*(t) )$ & nil \\
Case 2 & Level prem, with lapse support & 0 & Diff & $ - \pi(t) \, \theta \, ( \mu_{x+t}^{(2)} - \mu_{x+t} ) \, ( S - V^*(t) )$ & $ - \pi(t) \, \theta \, \nu_{x+t} \, V^*(t)$  \\
Case 3 & Premium = mortality cost & n/a & Unif & $ - \pi(t) \, \theta \, ( \mu_{x+t}^{(2)} - \mu_{x+t} ) \, S$ & nil \\
Case 3 & Premium = mortality cost & n/a & Diff & $ - \pi(t) \, \theta \, ( \mu_{x+t}^{(2)} - \mu_{x+t} ) \, S$ & nil
\end{tabular}
\end{center}
\end{table}
\end{landscape}

%-----------------------------------------------------------

\begin{bajlist}

\item Mortality surplus is similar in all cases, being weighted by $\pi(t)$ and differing only in the deduction of $V(t), V^*(t)$ or zero from the sum at risk. % (note that $V(t) \ge V^*(t) \ge 0$). We expect mortality surplus to be qualitatively similar in all cases.

\item In two cases, lapsing behavior (`Unif' {\em versus} `Diff') makes no difference, these are: level premiums, no lapse support and surrender values = $V(t)$; and premiums equal to mortality cost. EPV[losses] will still depend on lapse rates (see equation (\ref{eq:TotSurp})) but can be regarded as `pure' mortality losses.

\item Level premiums, no lapse support and nil surrender values stands apart from the other examples. Lapse surplus is positive, and always has a term weighted by $(1-\pi(t))$ {\em and} a term weighted by something close to lapse rate $\nu_{x+t}$. We expect surplus to be large and positive (consequently, relatively poor value for policyholders), with very little dependence on lapse behavior (consequently, small difficulty in managing lapse risk).

\item The main conclusion is that lapse-supported premiums do increase adverse selection losses, under differential lapsing (which it is reasonable to expect). How much of a difference this makes depends heavily on the weight $\pi(t) \, \theta \, \nu_{x+t}$, assuming that $C(t)=0$.

\end{bajlist}

\begin{table}
\begin{center}
\caption{\label{table:Losses} Examples of rates of adverse selection loss, per in-force policy, with proportional mortality hazards $\mu_{x+t}^{(2)} = 5 \, \mu_{x+t}^{(1)}$. Times $t$ such that $\pi(t)=0.001$ in all cases. Sum insured in `high-risk' subpopulation $10 \, S$. Valuation lapse rate $\nu_{x+t}=0.06$ and `normal' experience lapse rate $\nu_{x+t}^{(1)}=\nu_{x+t}=0.06$. `Unif' means uniform lapsing $\nu_{x+t}^{(2)} = \nu_{x+t}^{(1)}$ and `Diff' means differential lapsing $\nu_{x+t}^{(2)} = 0$.}
\small
\begin{tabular}{lllll}
& & & & \\
& & Surr & & \\
Case   & Description & Value & Lapsing & Rate of Adverse Selection Loss \\[0.5ex]
Case 1 & Level prem, no lapse support & 0      & Unif & $-0.04 \, \mu_{x+t}^{(1)} \, (S - V(t)) + 0.06054 \, V(t)$ \\
Case 1 & Level prem, no lapse support & 0      & Diff & $-0.04 \, \mu_{x+t}^{(1)} \, (S - V(t)) + 0.05994 \, V(t)$ \\
Case 1 & Level prem, no lapse support & $V(t)$ & Unif & $-0.04 \, \mu_{x+t}^{(1)} \, (S - V(t))$ \\
Case 1 & Level prem, no lapse support & $V(t)$ & Diff & $-0.04 \, \mu_{x+t}^{(1)} \, (S - V(t))$ \\
Case 2 & Level prem, with lapse support & 0 & Unif & $-0.04 \, \mu_{x+t}^{(1)} \, (S - V^*(t))$ \\
Case 2 & Level prem, with lapse support & 0 & Diff & $-0.04 \, \mu_{x+t}^{(1)} \, (S - V^*(t)) - 0.00060 \, V^*(t)$ \\
Case 3 & Premium = mortality cost & n/a & Unif & $-0.04 \, \mu_{x+t}^{(1)} \, S$ \\
Case 3 & Premium = mortality cost & n/a & Diff & $-0.04 \, \mu_{x+t}^{(1)} \, S$

\end{tabular}
\end{center}
\end{table}

%-------------------------------------------------------------

Table \ref{table:Losses} illustrates Table \ref{table:LossesTheory}, with the following choice of parameters: (a) an arbitrary mortality hazard $\mu_{x+t}^{(1)}$: (b) $\pi(t) = 0.001$; (c) mortality hazard in `high-risk' subpopulation $\mu_{x+t}^{(2)} = 5 \times \mu_{x+t}^{(1)}$; (d) basic lapse rate $\nu_{x+t}=0.06$; and (e) $\theta=10$. 

For simplicity we have illustrated each rate of loss at {\em some} time $t$ {such that} $\pi(t) = 0.001$, generally not the same time $t$ in each example. Table \ref{table:Losses} shows rates per in-force policy so does not allow for $\pi(t)$ changing over time, see Section \ref{sec:Attrition}. This is properly accounted for in Section \ref{sec:ThreeCasesEPV}. 

Table \ref{table:Losses} illustrates the main points: that the mortality surpluses are all of a kind; and lapse surplus with no lapse-support is positive and orders of magnitude larger than the losses arising from lapses with lapse-support. Numerical examples are in Section \ref{sec:NumExAdvSel}.

\subsection{Second Moments}
\label{sec:SecondMoments}

It is straightforward to write down the second moments of insurance losses, by partitioning the population according to mortality risk (see for example \cite{pollard1970}). Suppose the $j$th of $n$ sub-populations is homogeneous in respect of mortality and lapse risk, and for a given premium function $P(t)$ {let $\Gamma_j$ be a random variable with the same distribution as the loss random variable per individual in sub-population $j$, per unit sum insured}. Suppose the proportion in sub-population $j$ is $\pi_j$, and for the $i$th individual define the random variable $Y^i_j$ to be the indicator of presence in sub-population $j$, so $\Mean[Y^i_j] = \pi_j$. Let the sum insured taken out by insured persons in sub-population $j$ be $S_j$. {Finally, let $\Gamma^i$ be the loss random variable for the $i$th individual, and let $\Gamma$ be the loss random variable for an individual chosen at random}. Then:

\begin{equation}
{\Mean[\Gamma]} = \sum_j S_j \, \Prob[Y^i_j=1] \, \Mean[\Gamma^i \mid Y^i_j=1] = \sum_j \pi_j \, S_j \, \Mean[\Gamma_j]
\end{equation}

\noindent and:

\begin{equation}
{\Mean[(\Gamma)^2]} = \sum_j (S_j)^2 \, \Prob[Y^i_j=1] \, \Mean[(\Gamma^i)^2 \mid Y^i_j=1] = \sum_j \pi_j \, (S_j)^2 \, \Mean[(\Gamma_j)^2]
\end{equation}

\noindent noting that $(Y^i_j)^2 = Y^i_j$ and $Y^i_j \, Y^i_k = 0$ for $j \not= k$. Therefore:

\begin{eqnarray}
{\Var[\Gamma]} & = & \sum_j \pi_j \, (S_j)^2 \, \Mean[(\Gamma_j)^2] - \sum_j \sum_k \pi_j \, \pi_k \, S_j \, S_k \, \Mean[\Gamma_j] \, \Mean[\Gamma_k] \nonumber \\
& = & \sum_j \pi_j \, (1 - \pi_j) \, (S_j)^2 \, \Mean[(\Gamma_j)^2] + \sum_j (\pi_j)^2 \, (S_j)^2 \, \big( \Mean[(\Gamma_j)^2] - (\Mean[\Gamma_j])^2  \big) \nonumber \\
& & \quad - 2 \sum_{j \not= k} \pi_j \, \pi_k \, S_j \, S_k \, \Mean[\Gamma_j] \, \Mean[\Gamma_k]. \label{eq:SecondMoment}
\end{eqnarray}

In the special case of two subpopulations, high-risk with prevalence $p$, sum insured $S_H$ and unit loss $\Gamma_H$ and low-risk with prevalence $(1-p)$, sum insured $S_L$ and unit loss $\Gamma_L$, equation (\ref{eq:SecondMoment}) gives an overall variance of loss of:

\begin{equation}
{\Var[\Gamma]} = (1-p) \, S_L^2 \, \Mean[(\Gamma_L)^2] + p \, S_H^2 \, \Mean[(\Gamma_H)^2] - \big( (1-p) \, S_L \, \Mean[\Gamma_L] + p \, S_H \, \Mean[\Gamma_H] \big)^2. \label{eq:SecondMoment2}
\end{equation}

\noindent Numerical examples are in Section \ref{sec:NumExAdvSel}.

\newpage

%-------------------------------------------------------------
%-------------------------------------------------------------
%-------------------------------------------------------------

\section{Numerical Examples of Adverse Selection Costs}
\label{sec:NumExAdvSel}

\subsection{A Measure of Adverse Selection Costs}
\label{sec:Measure}

The measure of adverse selection costs we use is the following:

\begin{equation}
\mbox{Adverse selection cost} = 100 \times \frac{\mbox{EPV[Adverse selection losses]}}{\mbox{EPV[Premiums]}} \, \%. \label{eq:Measure}
\end{equation}

\noindent This measure has several advantages. (a) it is scale-free; (b) it can be interpreted simply as {(minus)} the overall percentage increase in premiums necessary to pay for the cost of adverse selection; (c) it automatically adjusts for additional income due to adverse selection as well as additional cost (individuals who take out larger sums insured also must pay higher premiums); and (d) it has been used in most studies of genetic testing and adverse selection that we consider in this paper\footnote{This measure is used in \cite{macdonald2011} and all its {antecedent} works, and it is one measure used in \cite{howard2014}. \cite{lombardo2018} projected future outgo under adverse selection but not future income.}.

%------------------------------------------------------------

\begin{table}
\begin{center}
\caption{\label{table:AdvSel1} Adverse selection costs as percentage of EPV[Premiums] with $\mu_{x+t}^{(2)} = \phi \mu_{x+t}^{(1)}$. Non-participating whole life contract endowing at age 100, taken out at age 35. Starting proportion in `high-risk' subpopulation $\pi_0=0.001$ and sum insured $\theta S$. {Force of interest 0.05}, valuation lapse rate $\nu_{x+t}=0.06$ and `normal' experience lapse rate $\nu_{x+t}^{(1)}=\nu_{x+t}=0.06$. `Unif' means uniform lapsing $\nu_{x+t}^{(2)} = \nu_{x+t}^{(1)}$ and `Diff' means differential lapsing $\nu_{x+t}^{(2)} = 0$.}
\small
\begin{tabular}{lld{4.2}d{2.4}d{4.2}d{2.4}d{4.2}d{2.4}d{4.2}d{2.4}}
& & & & & & & & & \\
& & \multicolumn{8}{c}{Adverse Selection Costs as \% of EPV[Premiums]} \\[0.5ex]
&     & \multicolumn{2}{c}{Case 1, $C(t)=0$} & \multicolumn{2}{c}{Case 1, $C(t)=V(t)$} & \multicolumn{2}{c}{Case 2, $C(t)=0$} &  \multicolumn{2}{c}{Case 3, $C(t)=0$} \\
&     & \multicolumn{2}{c}{No Lapse Support} & \multicolumn{2}{c}{No Lapse Support} & \multicolumn{2}{c}{With Lapse Support} & \multicolumn{2}{c}{Prem = Mort Cost} \\[0.5ex]
$\phi$ & $\theta$ & \multicolumn{1}{c}{Unif} & \multicolumn{1}{c}{Diff} & \multicolumn{1}{c}{Unif} & \multicolumn{1}{c}{Diff} & \multicolumn{1}{c}{Unif} & \multicolumn{1}{c}{Diff} & \multicolumn{1}{c}{Unif} & \multicolumn{1}{c}{Diff} \\[0.5ex]              
&     & \multicolumn{1}{c}{(\%)} & \multicolumn{1}{c}{(\%)} & \multicolumn{1}{c}{(\%)} & \multicolumn{1}{c}{(\%)} & \multicolumn{1}{c}{(\%)} & \multicolumn{1}{c}{(\%)} & \multicolumn{1}{c}{(\%)} & \multicolumn{1}{c}{(\%)} \\[0.5ex] 
   1$\times$ &    1$\times$ &    51.58  &    51.48 &     0.00 &     0.00 &     0.00 &    -0.20 & 0.00 &    0.00 \\
   1$\times$ &    4$\times$ &    51.58  &    51.19 &     0.00 &     0.00 &     0.00 &    -0.80 & 0.00 &    0.00 \\
   1$\times$ &   10$\times$ &    51.58  &    50.62 &     0.00 &     0.00 &     0.00 &    -1.98 & 0.00 &    0.00 \\[0.5ex]
   2$\times$ &    1$\times$ &    51.54  &    51.40 &    -0.03 &    -0.09 &    -0.08 &    -0.37 & -0.09 &    -0.27 \\
   2$\times$ &    4$\times$ &    51.43  &    50.87 &    -0.14 &    -0.36 &    -0.32 &    -1.48 & -0.35 &    -1.08 \\
   2$\times$ &   10$\times$ &    51.20  &    49.81 &    -0.35 &    -0.89 &    -0.78 &    -3.65 & -0.87 &    -2.65 \\[0.5ex]
   5$\times$ &    1$\times$ &    51.46  &    51.26 &    -0.12 &    -0.25 &    -0.26 &    -0.67 & -0.28 &    -0.65 \\
   5$\times$ &    4$\times$ &    51.08  &    50.30 &    -0.47 &    -0.98 &    -1.04 &    -2.65 & -1.11 &    -2.58 \\
   5$\times$ &   10$\times$ &    50.33  &    48.40 &    -1.16 &    -2.44 &    -2.58 &    -6.57 & -2.77 &    -6.39
\end{tabular}
\end{center}
\end{table}

We are aware that this measure does not recognize the plausible response of consumers to higher prices, namely to reduce the amount of insurance purchased. However, that is a question we have to set aside for future research, {given that our measure should capture the first-order effects of adverse selection.}

\begin{figure}
\begin{center}
\includegraphics[scale=0.40]{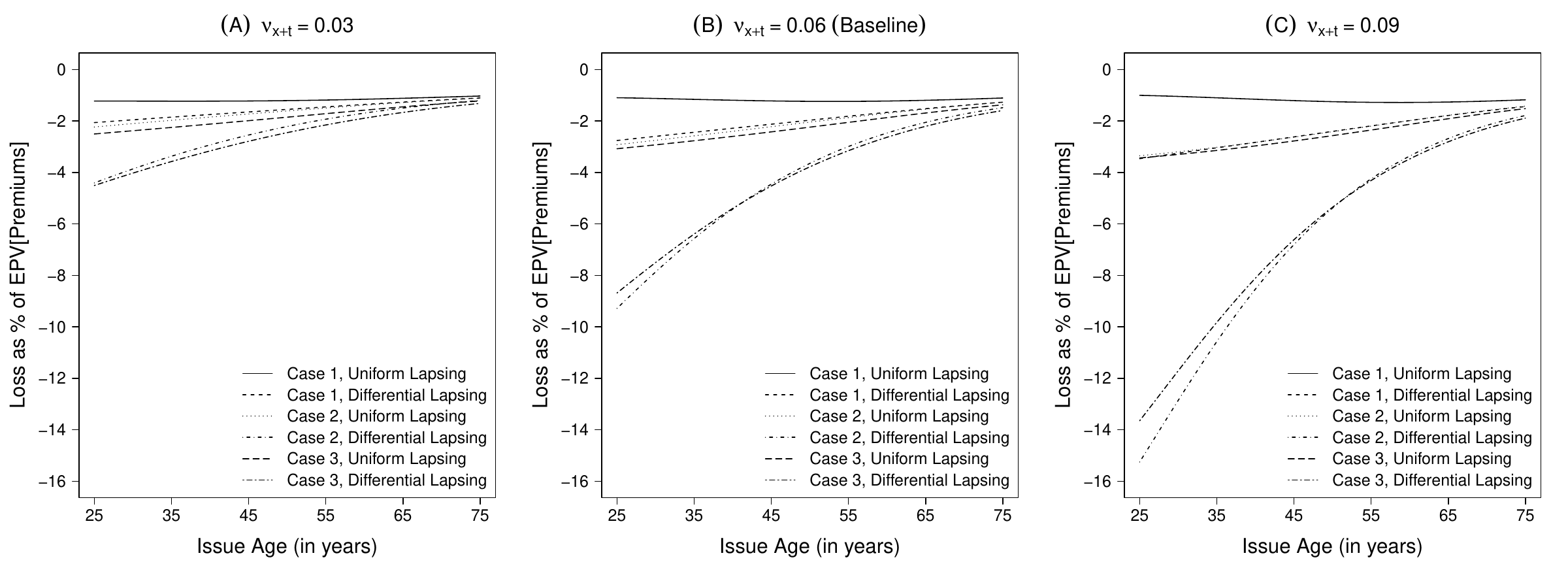}
\caption{\label{fig:losses} Adverse selection costs as percentage of EPV[Premiums] with $\mu_{x+t}^{(2)} = 5 \times  \mu_{x+t}^{(1)}$. Non-participating whole life contract endowing at age 100, taken out at ages {25--75}. Starting proportion in `high-risk' subpopulation $\pi_0=0.001$ and sum insured $10 \times S$. {Force of interest 0.05}. {Valuation lapse rate (A) $\nu_{x+t}=0.03$, (B) $\nu_{x+t}=0.06$ (baseline), and (C) $\nu_{x+t}=0.09$} and `normal' experience lapse rate $\nu_{x+t}^{(1)}=\nu_{x+t}=0.06$. `Uniform lapsing' means  $\nu_{x+t}^{(2)} = \nu_{x+t}^{(1)}$ and  `Differential lapsing' means $\nu_{x+t}^{(2)} = 0$.} 
\end{center}
\end{figure}

%-------------------------------------------------------------

\subsection{Numerical Examples of Adverse Selection Costs in Cases 1, 2 and 3}
\label{sec:ThreeCasesEPV}

We continue the examples from Section \ref{sec:Examples}, based on `Term to 100' contracts taken out at age 35, but now with a small `high-risk' subpopulation as in Section \ref{sec:ThieleInhom}, in a position to exercise adverse selection. 

% Our results are {not} intended to represent any real contract or market (in particular we ignore expenses), but to give insights into how lapse behavior affects adverse selection costs. The lapse rate of 0.06 per annum is chosen for comparison with real examples [REFERENCE TO BACK THIS UP] and the studies \cite{howard2014} and \cite{lombardo2018}, {see Appendix 2}. 

Table \ref{table:AdvSel1} shows adverse selection costs, as a percentage of EPV[Premiums] for selected values of $\phi$ (multiple of `normal' mortality) and $\theta$ (multiple of `normal' sum insured). {Figure \ref{fig:losses} shows how the losses vary with {entry ages from 25 to 75}, taking the most extreme scenario ($\phi=5, \theta=10$) from Table \ref{table:AdvSel1}.} {The figure shows three valuation lapse rates, $\nu_{x+t} = 0.03, 0.06$ (baseline) and $0.09$.} We consider separately the choice of basis, the main results and other observations. 

\subsubsection{Choice of Basis for Table \ref{table:AdvSel1} {and Figure \ref{fig:losses}}}
\label{sec:TableBasis}

The basis follows that of Table \ref{table:Losses}. We fix {force of interest $\delta=0.05$}, $\pi_0=0.001$  and $\nu_{x+t}^{(1)}=0.06$, just by way of example\footnote{The population prevalences of the thirteen disorders considered in \cite{howard2014} and \cite{lombardo2018} ranged from 1 in 500 to 1 in 20,000. Recall that $\pi_0$ is the starting proportion of lives in the `high-risk' sub-population.}\textsuperscript{,}\footnote{\cite{lombardo2018} cites historic lapse rates of 6.3\% per year in the United States. \cite{howard2014} assumed `normal' lapse rates of 3\% per year in Canada.}. `Normal' mortality was {GM82 Males}, and `high-risk' mortality was given by a {\em force} of mortality $\phi$ times `normal' ($\phi = 1, 2$ and 5). `High-risk' sums insured were $\theta$ times `normal', $\theta=1$ representing `precautionary adverse selection' and $\theta = 4, 10$ representing `speculative adverse selection'. There were no expenses.

\subsubsection{Main Results From Table \ref{table:AdvSel1} {and Figure \ref{fig:losses}}}
\label{sec:TableResults}

Recall from Section \ref{sec:IntroPlan} that our first main question was whether lapse-supported premiums affected an insurer's exposure to adverse selection risk. Our second main question concerned the robustness of methods used to illustrate adverse selection costs, if insurers were banned from using genetic test results.

\begin{bajlist}

\item The table shows costs without lapse-support in the two extreme examples of Case 1, $C(t)=0$ and $C(t)=V(t)$. The latter shows losses very much smaller than those under Case 2 with lapse support. In addition we see that:

\begin{bajsublist}
\item without lapse-support, a modest margin in surrender values ($C(t)\le 0.9 V(t)$ for example) would more than compensate for any adverse selection losses; and
\item Case 2, with lapse support appears to generate losses of the order of 1\% of premiums even when $\phi=1$, and there is no extra mortality risk at all. However, in the absence of extra mortality there should be no reason for differential lapsing to occur.
\end{bajsublist}

\noindent Overall, the answer to the first question is `yes', lapse-supported premiums substantially increase the possible adverse selection losses, especially if there is `speculative adverse selection'. An insurer writing lapse-supported business may have good reason to feel uneasy about restrictions on underwriting, as well as the secondary market.

\item The second question, concerning models illustrating adverse selection costs, compares the models of \cite{howard2014} and \cite{lombardo2018}, who used level premiums with an indeterminate degree of lapse-support; and \cite{macdonald2011}, who used premiums equal to mortality cost, and no explicit lapse assumptions. The most relevant comparison in Table \ref{table:AdvSel1} is therefore {between the last two pairs of} columns.

%%%%%%%%%%%%%%%%%%%%%%%%%%%%%%%%%%%%%%%%%%%%%%%%%%%%%%%%%%%%%%

\begin{table}
\begin{center}
\caption{\label{table:SecondMoments1} Standard deviation of adverse selection costs as proportion of EPV[Premiums] with $\mu_{x+t}^{(2)} = \phi \mu_{x+t}^{(1)}$. Starting proportion in `high-risk' subpopulation $\pi_0=0.001$ and sum insured $\theta S$. {Force of interest 0.05}, valuation lapse rate $\nu_{x+t}=0.06$ and `normal' experience lapse rate $\nu_{x+t}^{(1)}=\nu_{x+t}=0.06$. `Unif' means uniform lapsing $\nu_{x+t}^{(2)} = \nu_{x+t}^{(1)}$ and `Diff' means differential lapsing $\nu_{x+t}^{(2)} = 0$.}
\small
\begin{tabular}{lld{4.2}d{2.4}d{4.2}d{2.4}d{4.2}d{2.4}d{4.2}d{2.4}}
& & & & & & & & & \\
& & \multicolumn{8}{c}{Standard Deviation of Adverse Selection Costs as Proportion of EPV[Premiums]} \\[0.5ex]
&     & \multicolumn{2}{c}{Case 1, $C(t)=0$} & \multicolumn{2}{c}{Case 1, $C(t)=V(t)$} & \multicolumn{2}{c}{Case 2, $C(t)=0$} &  \multicolumn{2}{c}{Case 3, $C(t)=0$} \\
&     & \multicolumn{2}{c}{No Lapse Support} & \multicolumn{2}{c}{No Lapse Support} & \multicolumn{2}{c}{With Lapse Support} & \multicolumn{2}{c}{Prem = Mort Cost} \\[0.5ex]
$\phi$ & $\theta$ & \multicolumn{1}{c}{Unif} & \multicolumn{1}{c}{Diff} & \multicolumn{1}{c}{Unif} & \multicolumn{1}{c}{Diff} & \multicolumn{1}{c}{Unif} & \multicolumn{1}{c}{Diff} & \multicolumn{1}{c}{Unif} & \multicolumn{1}{c}{Diff} \\[0.5ex]              
   1$\times$ &    1$\times$ &     1.63  &     1.63 &     1.44 &     1.44 &     3.23 &     3.23 &     3.36 &     3.35 \\
   1$\times$ &    4$\times$ &     1.64  &     1.64 &     1.45 &     1.45 &     3.24 &     3.25 &     3.38 &     3.37 \\
   1$\times$ &   10$\times$ &     1.70  &     1.71 &     1.50 &     1.54 &     3.35 &     3.43 &     3.49 &     3.59 \\[0.5ex]
   2$\times$ &    1$\times$ &     1.63  &     1.63 &     1.44 &     1.44 &     3.23 &     3.23 &     3.36 &     3.36 \\
   2$\times$ &    4$\times$ &     1.65  &     1.65 &     1.46 &     1.47 &     3.26 &     3.29 &     3.40 &     3.41 \\
   2$\times$ &   10$\times$ &     1.75  &     1.78 &     1.56 &     1.62 &     3.48 &     3.64 &     3.61 &     3.75 \\[0.5ex]
   5$\times$ &    1$\times$ &     1.63  &     1.64 &     1.45 &     1.45 &     3.23 &     3.24 &     3.37 &     3.37 \\
   5$\times$ &    4$\times$ &     1.67  &     1.69 &     1.49 &     1.51 &     3.32 &     3.38 &     3.46 &     3.51 \\
   5$\times$ &   10$\times$ &     1.88  &     1.98 &     1.70 &     1.82 &     3.80 &     4.11 &     3.93 &     4.23 
\end{tabular}
\end{center}
\end{table}

%%%%%%%%%%%%%%%%%%%%%%%%%%%%%%%%%%%%%%%%%%%%%%%%%%%%%%%%%%%%%%

\begin{bajsublist}
\item Ignoring the results from Case 2, lapse-supported premiums,  $\phi=1$, for the same reasons as above, Case 2 and Case 3 give very similar results, for any practical purpose. Table \ref{table:LossesTheory} suggests why this is so: Case 2 has the smaller mortality loss, offset by a non-zero lapse loss. The exact balance will depend on circumstances.
\item Any reduction in the degree of lapse-support in Case 2 (for example, surrender values substantially greater than zero) would reduce adverse selection losses. There is no counterpart in Case 3.
\end{bajsublist}

\noindent Overall, the answer to the second question appears to be that the two methods are comparable in illustrating the extremes of adverse selection costs, but that the model with premiums equal to mortality cost is less flexible in representing lesser degrees of lapse-support. That is, it is generally more conservative. \cite{macdonald2011} suggested modeling a smaller insurance market as a proxy for lapses.

\item {Figure \ref{fig:losses} shows that differential lapsing is costly in all cases, and that losses drop significantly as age at entry increases, except for Case 1 with uniform lapsing. By age 75 at entry the losses in all cases have converged to a range of 1--2\%. Note that this figure shows the worst combination of mortality and lapsing from Table \ref{table:AdvSel1}.}

\end{bajlist}

{It appears that} whichever method we choose in a model of adverse selection losses, we will reach broadly similar conclusions. This tends to be borne out by such limited comparisions as can be made between \cite{macdonald2011} and \cite{howard2014}, see in particular the Appendix of the latter. Any difference between the conclusions of these models  probably cannot be attributed to lapse-supported premiums, if indeed these feature at all\footnote{The model premium rates used in \cite{howard2014} were an average of the actual rates charged by Canadian insurers. Policies were convertible into `Term to 100' policies at age 65. However, we noted in Section \ref{sec:Examples} that we cannot tell from premium rates alone whether or not they are lapse-supported.}.

%%%%%%%%%%%%%%%%%%%%%%%%%%%%%%%%%%%%%%%%%%%%%%%%%%%%%%%%%%%%%%

\subsubsection{Second Moments of Losses}
\label{sec:SecondMomentsExamples}

Table \ref{table:SecondMoments1} shows standard deviations of insurance losses (see Section \ref{sec:SecondMoments}), expressed as a proportion (note: not percentage) of the EPV of premiums, for the same values of higher mortality (factor $\phi$) {and higher sums insured under speculative adverse selection (factor $\theta$.) The clear conclusions are the following.

\begin{bajlist}

\item Standard deviations in Cases 2 and 3 are very close to each other, and both approximately double those in Case 1 (no lapse-support). 

\item Higher mortality (choice of $\phi$) and policyholder behavior (choice of $\theta$, uniform or differential lapsing) make very little difference within any benefits regime (Case 1, 2 or 3), except perhaps at the extreme ($\phi=10$).  

\end{bajlist}

} % END RED

\noindent It is possible, of course, that the distribution of insurance losses is poorly described by a conventional second moment. 

\newpage

%%%%%%%%%%%%%%%%%%%%%%%%%%%%%%%%%%%%%%%%%%%%%%%%%%%%%%%%%%%%%%

\subsubsection{Other Observations}
\label{sec:Observations}

\begin{bajlist}

\item In every case where losses are shown (negative entries) the losses are almost exactly in proportion to $\theta$, the `high-risk' multiple of the `normal' sum insured. Bearing this in mind, and following \cite{macdonald2011}, it suffices to report losses for average sums insured ($\theta=1$), {as we do in Table \ref{table:SecondMoments1}.}

% This supports statements in \cite{macdonald2011} and references therein, that it suffices to report losses for average sums insured, and infer further losses from `speculative adverse selection'.

\item {The choice of a contract extending cover to age 100 may seem extreme, but we note that \cite{howard2014} and \cite{lombardo2018} were based on contracts with benefits extending to at least age 100.}

\end{bajlist}

\subsection{Lapse-supported Premiums are Different}
\label{sec:LSPDifferent}

One important difference is that lapse-supported premiums are sensitive to any difference between the experienced lapse rate and the lapse rate assumed in the premium basis; the other two methods are not. If it is true that policies sold into the secondary market are less likely subsequently to lapse, it is logical for primary insurers to regard the secondary market as a threat, but that is not primarily because of adverse selection {\em at the point of sale}\footnote{One could regard the selection of policies to purchase by life settlement companies as a form of adverse selection exercised against the primary insurer after the point of sale.} whether associated with genetic testing or otherwise; it is simply in the nature of lapse-supported premiums.

\begin{table}
\begin{center}
\caption{\label{table:AdvSelSens} Sensitivity of adverse selection costs (percentage of EPV[Premiums]) to lapse rate in `normal' subpopulation. Non-participating whole life contract endowing at age 100, taken out at age 35. Proportional mortality hazards  $\mu_{x+t}^{(2)} = \phi \mu_{x+t}^{(1)}$. Starting proportion in `high-risk' subpopulation $\pi_0=0.001$ and sum insured $S$ (constant $\theta=1$). {Force of interest 0.05}, lapse rate 0.06 in premium/valuation basis, experience lapse rate $\tilde{\nu}_{x+t}^{(1)}$ in `normal' subpopulation. `Unif' means uniform lapsing $\nu_{x+t}^{(2)} = \nu_{x+t}^{(1)}$ and `Diff' means differential lapsing $\nu_{x+t}^{(2)} = 0$.}
\small
\begin{tabular}{lld{4.2}d{2.4}d{4.2}d{2.4}d{4.2}d{2.4}}
& & & & & & & \\
& & \multicolumn{6}{c}{Adverse Selection Costs as \% of EPV[Premiums]} \\[0.5ex]
&     & \multicolumn{2}{c}{Case 1, $C(t)=V(t)$} & \multicolumn{2}{c}{Case 2, $C(t)=0$} &  \multicolumn{2}{c}{Case 3, $C(t)=0$} \\
&     & \multicolumn{2}{c}{No Lapse Support} & \multicolumn{2}{c}{With Lapse Support} & \multicolumn{2}{c}{Prem = Mort Cost} \\[0.5ex]
$\phi$ & $\tilde{\nu}_{x+t}^{(1)}$ & \multicolumn{1}{c}{Unif} & \multicolumn{1}{c}{Diff} & \multicolumn{1}{c}{Unif} & \multicolumn{1}{c}{Diff} & \multicolumn{1}{c}{Unif} & \multicolumn{1}{c}{Diff} \\[0.5ex]
&      & \multicolumn{1}{c}{(\%)}   & \multicolumn{1}{c}{(\%)} & \multicolumn{1}{c}{(\%)} & \multicolumn{1}{c}{(\%)} & \multicolumn{1}{c}{(\%)} & \multicolumn{1}{c}{(\%)} \\[0.5ex] 
1$\times$ &   0.05 &     0.00 &     0.00 & -9.56 &    -9.72 &     0.00 &     0.00 \\
1$\times$ &   0.06 &     0.00 &     0.00 & 0.00 &    -0.20 &     0.00 &     0.00 \\
1$\times$ &   0.07 &     0.00 &     0.00 & 7.74 &     7.51 &     0.00 &     0.00 \\[0.5ex]
2$\times$ &   0.05 &    -0.04 &    -0.08 &    -9.64 &    -9.88 &    -0.09 & -0.23 \\
2$\times$ &   0.06 &    -0.03 &    -0.09 &    -0.08 &    -0.37 &    -0.09 & -0.27 \\
2$\times$ &   0.07 &    -0.03 &    -0.10 &     7.67 &     7.33 &    -0.09 & -0.32 \\[0.5ex]
5$\times$ &   0.05 &    -0.12 &    -0.23 &    -9.82 &   -10.15 &    -0.26 & -0.54 \\
5$\times$ &   0.06 &    -0.12 &    -0.25 &    -0.26 &    -0.67 &    -0.28 & -0.65  \\
5$\times$ &   0.07 &    -0.11 &    -0.27 &     7.49 &     7.01 &    -0.29 & -0.76 
\end{tabular}
\end{center}
\end{table}

\begin{table}
\begin{center}
\caption{\label{table:LossesSens} Examples of rates of adverse selection loss, per in-force policy, with proportional mortality hazards $\mu_{x+t}^{(2)} = 5 \, \mu_{x+t}^{(1)}$. Times $t$ such that $\pi(t)=0.001$ in all cases. Sum insured in `high-risk' subpopulation $10 \, S$. Valuation lapse rate $\nu_{x+t}=0.06$ and `normal' experience lapse rate $\tilde{\nu}_{x+t}^{(1)}=0.05$. `Unif' means uniform lapsing $\nu_{x+t}^{(2)} = \tilde{\nu}_{x+t}^{(1)}$ and `Diff' means differential lapsing $\nu_{x+t}^{(2)} = 0$.}
\small
\begin{tabular}{lllll}
& & & & \\
& & Surr & & \\
Case   & Description & Value & Lapsing & Rate of Adverse Selection Loss \\[0.5ex]
Case 1 & Level prem, no lapse support & 0      & Unif & $-0.04 \, \mu_{x+t}^{(1)} \, (S - V(t)) + 0.05045 \, V(t)$ \\
Case 1 & Level prem, no lapse support & 0      & Diff & $-0.04 \, \mu_{x+t}^{(1)} \, (S - V(t)) + 0.04995 \, V(t)$ \\
Case 1 & Level prem, no lapse support & $V(t)$ & Unif & $-0.04 \, \mu_{x+t}^{(1)} \, (S - V(t))$ \\
Case 1 & Level prem, no lapse support & $V(t)$ & Diff & $-0.04 \, \mu_{x+t}^{(1)} \, (S - V(t))$ \\
Case 2 & Level prem, with lapse support & 0 & Unif & $-0.04 \, \mu_{x+t}^{(1)} \, (S - V^*(t)) - 0.01009 \, V^*(t)$ \\
Case 2 & Level prem, with lapse support & 0 & Diff & $-0.04 \, \mu_{x+t}^{(1)} \, (S - V^*(t)) - 0.01059 \, V^*(t)$ \\
Case 3 & Premium = mortality cost & 0 & Unif & $-0.04 \, \mu_{x+t}^{(1)} \, S$ \\
Case 3 & Premium = mortality cost & 0 & Diff & $-0.04 \, \mu_{x+t}^{(1)} \, S$
\end{tabular}
\end{center}
\end{table}

% It would seem that lapse-supported premiums is just one `natural' way to dispose of lapse surplus, and in terms of adverse selection costs not substantially different from the alternative methods considered. However, it is the only method considered that depends on any element of the premium basis, in this case the lapse rate, being set correctly. Its true nature is shown more clearly if we test the three `natural' methods against a different lapse experience. 

The basic lapse rate in Table \ref{table:AdvSel1} was 6\% {\em per annum}, meaning that this rate was used in {lapse-supported} premium and valuation bases. In Table \ref{table:AdvSelSens} we show adverse selection costs (EPV[Premiums]) with experienced lapse rates, denoted by $\tilde{\nu}_{x+t}^{(1)}$, of {0.05, 0.06 and 0.07} {\em per annum}, and `high-risk' force of mortality of 1, 2 and 5 times `normal'. Now lapse-supported premiums are very different from the other two approaches. Under the latter, `stressing' the lapse rates by 1\% changes the profit/loss by very little --- in the worst case about 0.2\%. Under lapse-supported premiums the change is of the order of 10--15\%. To see why, modify the reasoning in Section \ref{sec:SurplusAdverse}. Now $\tilde{\nu}_{x+t}^{(1)} = 0.05 < 0.06 = {\nu}_{x+t}$, so the  `normal' subpopulation contributes lapse surplus at rate:

\begin{equation}
- (1-\pi(t)) \, (\tilde{\nu}_{x+t}^{(1)} - \nu_{x+t}) (C(t) - V^*(t)) \label{eq:Sens1}
\end{equation}

\noindent where before it contributed none, and this dominates the {rate of emerging surplus} in the `high-risk' subpopulation of:

\begin{equation}
- \pi(t) \, \theta \, \big( (\mu_{x+t}^{(2)} - \mu_{x+t})(S - V^*(t)) + ({\nu}_{x+t}^{(2)} - \nu_{x+t}) (C(t) - V^*(t))  \big) \label{eq:Sens2}
\end{equation} 

\noindent under uniform or differential lapsing. Table \ref{table:LossesSens} shows the emerging rates of adverse selection losses with $\tilde{\nu}_{x+t}^{(1)}=0.05$ (continuous-time cashflows). The main difference from Table \ref{table:Losses} is that Case 2 now generates similar losses under both uniform and differential lapsing.
 
\newpage

%----------------------------------------------------------------
%----------------------------------------------------------------
%----------------------------------------------------------------

\section{Conclusions}
\label{sec:Conclusions}

\subsection{Motivation: {Main Questions}}
\label{sec:Motivation}

% Recent studies of adverse selection in life insurance, arising because of genetic testing (\cite{howard2014}, \cite{lombardo2018}), have drawn attention (or caused it to be drawn) to two additional sources of risk.

% \begin{bajlist}
% \item Life settlement companies initiating the purchase of large sums insured by high-risk individuals. This was discussed at length in \cite{hacariz2020b}.
% \item Insurance purchases by high-risk individuals being concentrated on products with lapse-supported premiums, which are vulnerable to low lapse rates. That is the subject of this paper.
% \end{bajlist}

% It is possible that these risks mostly affect life insurance markets in North America.

{Our two main questions were: (a) how do lapse-supported premiums affect an insurer's exposure to the risk of adverse selection?; and (b) how robust are the conclusions of models used recently to illustrate the impact of banning insurers' access to genetic test results?}

% \subsection{Model}
% \label{sec:Model}

% We used a simple double-decrement model of death and lapses (Figure \ref{fig:Lapses}) in Section \ref{sec:Analysis} to write down Thiele's equations for premiums and policy values with and without lapse-support. We noted the classical origins of the model in \cite{lidstone1905} which anticipated modern {treatments} such as \cite{hoem1988}. 

% We used models in continuous time for theoretical development and models in discrete time for examples, each being best suited to its purpose, for which we claim some justification from Lidstone. Assuming that surrender values satisfied $C(t) = k \, V(t)$ led to equation (\ref{eq:Thiele3}), effectively known to Lidstone, which showed lapse-supported premiums to be equivalent to a change of interest rate. No such result seems to exist for models with discrete cashflows.

% However, equation (\ref{eq:Thiele3}) is a result on first moments only, meaning that it may be directly relevant for pricing  and reserving under traditional equivalence principles, but not for risk management. For risk management we need to know about higher moments of losses, and then losses arising from asset returns and lapses are not equivalent.

% In Section \ref{sec:ActModelLSP} we obtained Thiele's equations for the rate of adverse selection {surplus or} loss emerging in an inhomogeneous population, containing a small high-risk subpopulation of `adverse selectors'. These took a very simple and intuitive form under the continuous-time cashflow model, not {easily} reproducible under the discrete-time cashflow model. 

\subsection{Main results}
\label{sec:ConcsResults}

We identified three approaches to disposing of lapse surplus: 

\begin{bajlist}
\item retrospectively (Case 1: level premiums without lapse-support);
\item prospectively (Case 2: level premiums with lapse-support); and
\item eliminating it by policy design (Case 3: premiums equal to mortality cost).
\end{bajlist}

\noindent Our main examples, based on `Term to 100' contracts and comparing EPV[adverse selection losses], were in Table \ref{table:AdvSel1}.

% Our examples assumed a `normal' lapse rate of 6\% per year, which is consistent with actual experience in the US. They were not, however, intended to reproduce particular premium or valuation bases in use anywhere, the aim was simply to illustrate the dynamics of lapse-supported premiums.

\begin{bajlist}

\item Comparing Case 1 and Case 2 answered our first question: lapse-supported premiums increase significantly the costs of adverse selection. Without lapse support, lapse surpluses can provide a margin large enough to absorb such losses. % However this conclusion depends on the degree of lapse-support, which need not always be so extreme as in the canonical `Term to 100' contract. 

\item Comparing Case 2 and Case 3 answered our second question: published models of adverse selection costs arising from restrictions on insurers' access to genetic test results appear to be equally robust, given their different methodologies. Broadly, \cite{howard2014} and \cite{lombardo2018} fall under Case 2, while \cite{macdonald2011} falls under Case 3. Table \ref{table:LossesTheory} explained this observation by comparing rates of earned surplus (losses). 

\item The underlying reason in both cases was that lapse-supported premiums (Case 2) were strongly sensitive to differences between experienced lapse rates and the lapse rates assumed in the premium basis, while the other methodologies were not. Tables \ref{table:AdvSelSens} and \ref{table:LossesSens} showed why this is so, and that it is a property of lapse-supported premiums as such, irrespective of adverse selection.

\item {Second moments of insurance losses showed striking results: {sensitivity to the high-level premium regime; almost complete insensitivity} to policyholder behaviour within a given regime.}

\end{bajlist}

%------------------------------------------------------------%

\subsection{Other Comments}
\label{sec:ConcsOther}

% Our other observations are as follows.

\begin{bajlist}

\item It is not possible to tell from premium rates alone whether or not a contract is lapse-supported.

\item We noted a paucity of actuarial literature about lapse-supported premiums and life settlement companies, despite the practical challenges they pose to insurers and regulators. Most of the literature {on lapsation} is economic or econometric in nature.

\item We note, as others have, the conflict of interest faced by an insurer selling lapse-supported business: the business is more profitable if persistency is poor. 

\item Adverse selection losses were very nearly proportionate to: (i) the `high-risk' mortality hazard, as a proportion of the `normal' mortality hazard; and (ii) the sum insured chosen by `adverse selectors', as a proportion of the `normal' sum insured. All useful information was gained from a model of `precautionary adverse selection'.

\item The lapse-supported business model may be threatened by life settlement companies, who would aim to buy those policies least profitable to the insurer and keep them in force or, worse, initiate their purchase. We refer to \cite{hacariz2020b} for a full discussion.

\item We suggest that lapse-supported premiums {should be regarded as belonging} to that class of financial schemes which recognize profit not yet earned, though in a form other than cash.

\end{bajlist}

\acknowledgements

This study is part of the research programme at the Research Centre for Longevity Risk --- a joint initiative of NN Group and the University of Amsterdam, with additional funding from the Dutch Government's Public Private Partnership programme. {We are grateful to Dr Stephen Richards for comments on a draft of this paper.} The MSc thesis by Yifan Zhang at Heriot-Watt University in summer 2022 was helpful in formulating the numerical examples.

\competinginterests

None.

%----------------------------------------------------------------
%----------------------------------------------------------------
%----------------------------------------------------------------

\bigskip
\begin{center}
{\sc Appendix 1}
\end{center}

\begin{appendix}

\noindent {\sc Measures of the Cost of Adverse Selection}

\medskip

In this Appendix we show the rates at which surplus emerges in a heterogeneous population with two subpopulationss, see Section \ref{sec:SurplusAdverse}.

\medskip
\noindent {\it Case 1: Level Premiums, No Lapse Support}
\medskip

Let $W(t)$ be the rate at which surplus is earned at time $t$ per policy in force. By the same reasoning as in Section \ref{sec:LapseSurplus} applied separately to: (a) a policy in the `normal' subpopulation with sum insured $S$ and; (b) a policy in the `high-risk' subpopulation with sum insured $\theta S$:

\begin{eqnarray}
W(t) & = & - (1 - \pi(t)) \left[ (\mu_{x+t}^{(1)} - \mu_{x+t}) (S - V(t)) + (\nu_{x+t}^{(1)} - \nu_{x+t}) (C(t) - V(t)) \right] \nonumber \\
& & - \pi(t) \, \theta \left[ (\mu_{x+t}^{(2)} - \mu_{x+t}) (S - V(t)) + (\nu_{x+t}^{(2)} - \nu_{x+t}) (C(t) - V(t)) \right]. \label{eq:dW1}
\end{eqnarray}

\noindent Since $\nu_{x+t}=0$ and we assume $\mu_{x+t}^{(1)}=\mu_{x+t}$ this simplifies to:

\begin{eqnarray}
W(t) & = & - \pi(t) \, \theta \, ( \mu_{x+t}^{(2)} - \mu_{x+t} ) (S - V(t)) \nonumber \\
&   & \quad + \Big( (1-\pi(t)) \, \nu_{x+t}^{(1)} + \pi(t) \, \theta \, \nu_{x+t}^{(2)} \Big) (V(t) - C(t))
\label{eq:dW1b}
\end{eqnarray}

\noindent expressed as mortality and lapse surplus components, of opposite sign because $\mu_{x+t}^{(2)} \ge \mu_{x+t}$ and $S \ge V(t) \ge C(t)$. Any positive lapse rates will generate positive surplus, while increased mortality $\mu_{x+t}^{(2)}$ in the `high-risk' subpopulation will generate negative surplus.

% We assume that lapse surplus may fund some level of surrender values, between the two extremes of no distribution ($C(t)=0$) and full distribution ($C(t) = V(t)$). If surrender values are guaranteed, we suppose that the guarantees take the form of a proportion of the policy value.

\medskip
\noindent{\it Case 2: Level Premiums, With Lapse Support}
\medskip

% Let $C^*(t)$ be the surrender values with lapse support, and let $V^*(t)$ and $P^*$ be policy values and level premium rate satisfying the following version of Thiele's equation:  

% \begin{equation}
% \frac{d}{dt} V^*(t) = \delta \, V^*(t) + P^* - \mu_{x+t} \, (S - V^*(t)) - \nu_{x+t} \, (C^*(t) - V^*(t))   \label{eq:Thiele2Rep}
% \end{equation}

% \noindent with boundary conditions $V^*(0)=0, V^*(n)= S$. 

{By similar reasoning to that above, the rate $W(t)$ at which surplus is earned per policy in force at time $t$ is:}

\begin{eqnarray}
W(t) & = & - (1 - \pi(t)) \left[ (\mu_{x+t}^{(1)} - \mu_{x+t}) (S - V^*(t)) + (\nu_{x+t}^{(1)} - \nu_{x+t}) (C(t) - V^*(t)) \right] \nonumber \\
& & - \pi(t) \, \theta \left[ (\mu_{x+t}^{(2)} - \mu_{x+t}) (S - V^*(t)) + (\nu_{x+t}^{(2)} - \nu_{x+t}) (C(t) - V^*(t)) \right]. \label{eq:ThieleLSPRep} 
\end{eqnarray}

\noindent If $\mu_{x+t}^{(1)} = \mu_{x+t}$ and $\nu_{x+t}^{(1)} = \nu_{x+t}$, the `normal' subpopulation generates no surplus, and:

\begin{equation}
W(t) = \pi(t) \, \theta \, \Big( - ( \mu_{x+t}^{(2)} - \mu_{x+t} ) (S - V^*(t)) - ( \nu_{x+t}^{(2)} -\nu_{x+t} ) \, (C(t) - V^*(t)) \Big). \label{eq:SurplusLSP1}
\end{equation} 

\noindent The two terms into which the right-hand side of equation (\ref{eq:SurplusLSP1}) splits are easily identified as the rates of accumulation of mortality surplus and lapse surplus, respectively. (Compare with the `critical function' components of \cite{lidstone1905}.)

\medskip
\noindent {\it Case 3: Premiums Equal to Mortality Cost}
\medskip

The premium function $P(t) = \mu_{x+t} (S-V(t))$ was defined in Section \ref{sec:PremFunc}, equation (\ref{eq:MortCost}). We noted there that as a consequence, $V(t)=0$ for all $t \ge 0$, hence $C(t)=0$ also. So all lapse surplus vanishes, and  the rate $W(t)$ at which surplus is earned is:

\begin{equation}
W(t) =  -(1 - \pi(t)) \, (\mu_{x+t}^{(1)} - \mu_{x+t}) \, S  - \pi(t) \, \theta \, (\mu_{x+t}^{(2)} - \mu_{x+t}) S.  
\end{equation}

\noindent If $\mu_{x+t}^{(1)} = \mu_{x+t}$ then only the second term remains. It eliminates lapse surplus, and any effects of lapse-supported premiums, {by design of the policy}. It therefore provides a `pure' benchmark for measuring adverse selection mortality costs, not contaminated by the treatment of lapses. Also, it is conservative, see Table \ref{table:Losses} and Section \ref{sec:SurplusAdverse}.

\end{appendix}

%----------------------------------------------------------------

\bigskip
\begin{center}
{\sc Appendix 2}
\end{center}

% \bigskip
% \begin{center}
% {\sc Appendices}
% \end{center}

\begin{appendix}

\noindent {\sc Genetic Testing and Adverse Selection in Life Insurance}

% \section{Genetic Testing and Adverse Selection in Life Insurance}

Warnings were first raised, that genetic testing could lead to problems in life and health insurance, in the late 1980s and early 1990s {\citep{pokorski1995}}. There was much pressure for genetic information of any kind to become a `protected characteristic' like race, sex and disability. In many jurisdictions, these characteristics (or certain of them) were not protected from insurance underwriting, meaning that different premiums could be charged based upon them, because exemptions had been inserted in relevant `anti-discrimination' laws. Insurers naturally pressed for similar exemptions to apply to any genetic `anti-discrimination' legislation. Lined up against them were advocacy groups and patient associations who evoked considerable public sympathy. Insurers possibly underestimated just how sensitive a topic genetics would be.

It would not be appropriate here to attempt a history of the arguments that ensued, or their consequences. Enough to say that it is not over yet, a very recent example being the Genetic Non-discrimination Act in Canada {\citep{prince2019}}: argued over in consultation, then in the Senate, then challenged in the federal courts after being enacted (it is now confirmed law).

Adverse selection has featured strongly in such debates, being perhaps the industry's main defence against arbitrary restrictions. But while insurers have had no problem in making the {\em theoretical} case that adverse selection is bad for them, they have been strikingly unable to produce much convincing {\em evidence}, of a quantitative nature, that adverse selection caused by  genetic information will pose a serious threat. Focussing on studies based on real genetic disorders and their epidemiology, we highlight the following lines of research, which helped to motivate this paper.

\begin{bajlist}

\item A programme of research based on modeling individual disorder, launched at Heriot-Watt University in 1999, led in time to \cite{macdonald2011}, which aggregated the impact of six representative disorders. The contracts had premiums equal to mortality cost (Section \ref{sec:PremFunc}), expiring at age 60, and {therefore} were not lapse-supported. It found adverse selection costs to be very small, a fraction of 1\% of premiums, except in certain extreme circumstances, because: (i) the disorders were very rare; (ii) {the main mortality risk was at pre-retirement ages; and (iii)} the work concentrated on `precautionary adverse selection' and did not highlight `speculative adverse selection' (see Section \ref{sec:LapseRisk}).

\item While Canadian Bill-201 was being promoted in the Canadian Senate, the Canadian Institute of Actuaries (CIA) commissioned a report \citep{howard2014} which found adverse selection costs to be significant, about 12\% of premiums. This report featured thirteen disorders, and highlighted `speculative adverse selection' by assuming that `adverse selectors' would take out ten times the normal amount of life insurance. The contracts were convertible term to age 65, convertible then to a `Term to 100' policy, with a conversion rate varying from nil (`standard' lives without Alzheimer's) to 100\% (`substandard' lives with Alzheimer's), and annual premiums obtained by reference to an industry database. Using a different measure, insured lives mortality was predicted to increase by 36\% for males and 58\% for females.

\item Later, the Society of Actuaries published a report \citep{lombardo2018} which adapted the analysis of \cite{howard2014} to US circumstances, and found costs to be less than 1\% of claims costs at first, rising to over 5\% after 30 years. The initial assumption was that both an in-force baseline block and future new business were whole-of-life to age 100 contracts. A second scenario assumed the in-force baseline consisted of renewable T20 plans only. `Adverse selectors' took out sums insured more than four times the average.

\item {Howard suggested \cite[Appendix]{howard2014} that the results of his model would be similar to the results of \cite{macdonald2011}, had it been possible to perform a side-by-side comparison. The results here, and those of earlier papers on cardiomyopathies \citep{hacariz2021, hacariz2022} lead us to agree with that statement. We note that \cite{howard2014} and \cite{lombardo2018}, focussed exclusively on `speculative adverse selection', in which regard we {point to} the conclusions of \cite{hacariz2020b}}.

\end{bajlist}

\end{appendix}

\bigskip

\renewcommand{\section}[1]{\noindent}
% \addcontentsline{toc}{section}{References} 
\bibliography{Bibliography.bib}

% \renewcommand{\section}[1]{\noindent}
% \addcontentsline{toc}{section}{References} 
% \bibliography{Bibliography.bib}

\end{document}